\documentclass[letterpaper,man,floatsintext,natbib, 12pt]{article}  

\usepackage[margin=0.7in]{geometry}
\usepackage[american]{babel}
\usepackage{amsmath}
\usepackage{graphicx}
\usepackage{xcolor, colortbl}
\usepackage{setspace}
\usepackage{threeparttable}
\usepackage{endnotes}
\usepackage{xspace}
\usepackage{afterpage}
\usepackage{mathtools}
\usepackage{soul}
\usepackage{bm}
\usepackage{dsfont}
\usepackage{amssymb}
\usepackage{bbm}
\usepackage{lscape}
\usepackage[normalem]{ulem}
\usepackage{csquotes}
\usepackage{algorithm}
\usepackage{algpseudocode} 
\usepackage{amsthm}
\theoremstyle{remark}
\newtheorem{definition}{Definition}

\newcommand{\HT}[1]{\hypertarget{(#1)}{(#1)}}
\newcommand{\cond}{\, \big| \,}

\newcommand{\indep}{\ \rotatebox[origin=c]{90}{$\models$} \ }

\newcommand{\PreserveBackslash}[1]{\let\temp=\\#1\let\\=\temp}
\newcolumntype{C}[1]{>{\PreserveBackslash\centering}p{#1}}
\newcolumntype{L}[1]{>{\PreserveBackslash\raggedright}p{#1}}
\newcolumntype{R}[1]{>{\PreserveBackslash\raggedleft}p{#1}}

\makeatother

\RequirePackage[colorlinks,citecolor=black,urlcolor=black, linkcolor = black ]{hyperref}

\usepackage{ragged2e}
\usepackage{multirow}
\usepackage{subcaption}
\usepackage{amsmath,amssymb,xcolor}
\usepackage{pgf,tikz}
\usetikzlibrary{arrows,shapes.arrows,shapes.geometric,shapes.multipart, decorations.pathmorphing,positioning,shapes.swigs,shapes,decorations,arrows,calc,arrows.meta,fit,positioning}
\tikzset{
    -Latex,auto,node distance =1 cm and 1 cm,semithick ,
    state/.style ={ellipse, draw, minimum width = 0.7 cm},
    point/.style = {circle, draw, inner sep=0.04cm,fill,node contents={}},
    bidirected/.style={Latex-Latex,dashed},
    el/.style = {inner sep=2pt, align=left, sloped},
    styRectDef/.style = {rectangle, rounded corners, draw=black, inner xsep=6mm, inner sep=3mm}
}

\usepackage[style=apa,sortcites=true,sorting=nyt,backend=biber]{biblatex}
\DeclareLanguageMapping{american}{american-apa}
\usepackage{sectsty}
\usepackage{authblk}
\sectionfont{\fontsize{14}{15}\selectfont}

\usepackage{footmisc}
\renewcommand\footnotelayout{%
  \advance\leftskip 1cm
  \advance\rightskip 1.2cm
 } 
\usepackage[font=small, margin=0.4in]{caption}
\sectionfont{\fontsize{13}{15}\selectfont} 
\subsectionfont{\fontsize{13}{15}\selectfont} 
\usepackage{lipsum}

\makeatletter
\newcommand{\algorithmfootnote}[2][\footnotesize]{%
  \let\old@algocf@finish\@algocf@finish
  \def\@algocf@finish{\old@algocf@finish
    \leavevmode\rlap{\begin{minipage}{\linewidth}
    #1#2
    \end{minipage}}%
  }%
}

\begin{document}

\title{\large{Identifying Causes of Test Unfairness: Manipulability and Separability} \\~\\ }

\author[1]{\normalsize{Youmi Suk}}
\author[2]{Weicong Lyu}

\affil[1]{\small{Teachers College, Columbia University; ysuk@tc.columbia.edu}}
\affil[2]{\small{University of Macau; weiconglyu@um.edu.mo}}

\date{\ \ } 
\maketitle

\abstract{Differential item functioning (DIF) is a widely used statistical notion for identifying items that may disadvantage specific groups of test-takers. These groups are often defined by non-manipulable characteristics, e.g., gender, race/ethnicity, or English-language learner (ELL) status. While DIF can be framed as a causal fairness problem by treating group membership as the treatment variable, this invokes the long-standing controversy over the interpretation of causal effects for non-manipulable treatments. To better identify and interpret causal sources of DIF, this study leverages an interventionist approach using treatment decomposition proposed by  \textcite{robins2010alternative}. Under this framework, we can decompose a non-manipulable treatment into intervening variables. For example, ELL status can be decomposed into English lexical proficiency and instructional English comprehension, each of which influences the outcome through different causal pathways. We formally define separable DIF effects associated with these decomposed components, depending on the absence or presence of item impact, and provide causal identification strategies for each effect. We then apply the framework to biased test items in the SAT and Regents exams. We also provide formal detection methods using causal machine learning methods, namely causal forests and Bayesian additive regression trees, and demonstrate their performance through a simulation study and a real-world application. Finally, we discuss the implications of adopting interventionist approaches in educational testing practices.
\\

}

\setcounter{secnumdepth}{3}

\section{Introduction}

Differential item functioning \parencite[DIF;][]{thissen1988use, pine1977} is currently one of the most widely used notions in the testing community to describe (un)fairness in test items. An item is flagged as exhibiting DIF if it functions differently for test-takers from different demographic or behavioral groups (e.g., gender, race/ethnicity) after conditioning on their abilities \parencite{holland1993differential, zumbo2007three}. 
While DIF has traditionally been treated as a statistical notion depicting associations between items and test-taker characteristics in observed data, a causal framework for DIF has recently become available \parencite{suk_lyu_2026}. This framework allows researchers to raise counterfactual questions, such as what an individual's performance would have been had they belonged to a different group (e.g., focal versus reference groups). However, this approach faces a long-standing challenge in treating non-manipulable treatments (e.g., gender, race) as ``causal'' variables.\footnote{\Textcite{holland2003causation} argues that causal variables are those that can be manipulated as treatments, whereas attributes like race, gender, or age are not causal because they cannot be assigned to different levels of them by intervention.} The overall goal of this study is to propose a new causal framework for DIF that focuses on the intervenable components of non-manipulable treatments to identify the ``real'' causes of test unfairness.

The fundamental idea of interventionism traces back to \citeauthor{holland1986}'s (\citeyear{holland1986}) maxim, ``no causation without manipulation,'' and has since been elaborated in extensive discussions \parencite[e.g.,][]{robins2022interventionist, dawid2000causal, greiner2011causal, vanderweele2014causal}. From this perspective, an individual's background characteristics, such as gender, race, and test scores, are not causes in themselves, though specific aspects of these variables (e.g., names, experience, skills) can be manipulated \parencite{holland2003causation, bertrand2004are}. Hence, asking what would have happened to a female test-taker if she had been male is conceptually ambiguous. Instead, it is more rigorous to ask what would have happened to her if specific components (e.g., experience, skills) had been changed in a way that is typical of males. For example, as we will demonstrate in later sections, we can evaluate  potential performance if gender-specific exposure to sports had been changed instead of gender itself. In this sense, while the causal DIF framework proposed by \Textcite{suk_lyu_2026} introduces causal reasoning to test fairness, it still faces the classic challenge of non-manipulable treatments, as it requires an intervention on group membership and, under certain conditions, on ability as well. Attributing causality to a non-manipulable treatment or mediator is often confusing and unhelpful for identifying actionable causes of test unfairness.  

Recently, Robins and his colleagues \parencite{robins2010alternative, robins2022interventionist} have reformulated mediation analysis in terms of interventions on \emph{treatment components} rather than on mediators. In this approach, treatment $A$ is decomposed into multiple intervening variables (e.g., $A_D$ and $A_Y$) that operate on outcomes via distinct causal pathways: one that directly affects the outcome and the other that indirectly affects the outcome through a mediator. Then it estimates so-called \emph{separable effects} of each component. This formulation requires only that interventions on $A_D$ and $A_Y$ are substantively meaningful, which makes it straightforward to discuss with subject matter experts (e.g., educators) regarding feasible experimental treatments \parencite{robins2022interventionist}. In prior work, this idea has commonly been applied to manipulable or intervenable treatments. For example, cigarette can be decomposed into nicotine and non-nicotine components, and cancer therapy into inhibitors of cancer cell proliferation and pain relievers; then the separable effects of these components are estimated \parencite{robins2010alternative, stensrud2022separable}. Unfortunately, there is limited research on how treatment decomposition can be used to interpret and estimate the causal effects of manipulable descendants of ill-defined, non-manipulable treatments. One notable exception is \Textcite{wen2024causal}, which introduces a single intervening component of a (possibly ill-defined) treatment in the mediator pathway. However, their work is limited in our setting because the DIF effect lies on the direct path from treatment to outcome, and may operate via multiple intervening components.

The main goal of this paper is to propose a novel causal framework for assessing test unfairness at the item level using treatment decomposition and separable effects. Specifically, by leveraging causal graphs and potential outcomes, we define \emph{separable} DIF as the difference in item functioning between two intervenable hypothetical worlds. We also categorize separable DIF into two types---simple separable DIF and general separable DIF---depending on the absence or presence of \emph{impact} (which assumes that the ability distribution differs across groups). We discuss nonparametric identification strategies and demonstrate how our approach can be conceptually applied to items from the SAT math exam and the New York Regents math exam. We also propose detection methods using causal forests \parencite{wager2018estimation, athey2019generalized} and Bayesian additive regression trees \parencite[BART;][]{hill2011bayesian, chipman2010BART} and investigate their performance via a simulation study and a real-world application.

\section{Motivating Examples}\label{sec:mot_ex}

\begin{figure}[!h]
    \centering
    \subcaptionbox{An SAT math item \label{fig:SAT}}[0.8\textwidth]  
    {    
    \includegraphics[width=0.7\linewidth]{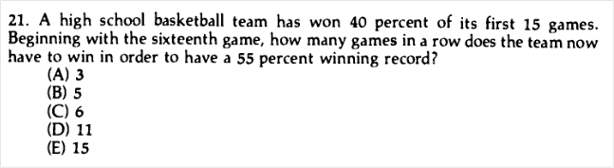}}
     \subcaptionbox{A Regents math item  \label{fig:regent}}[.8\textwidth]  
    {
        \includegraphics[width=0.7\linewidth]{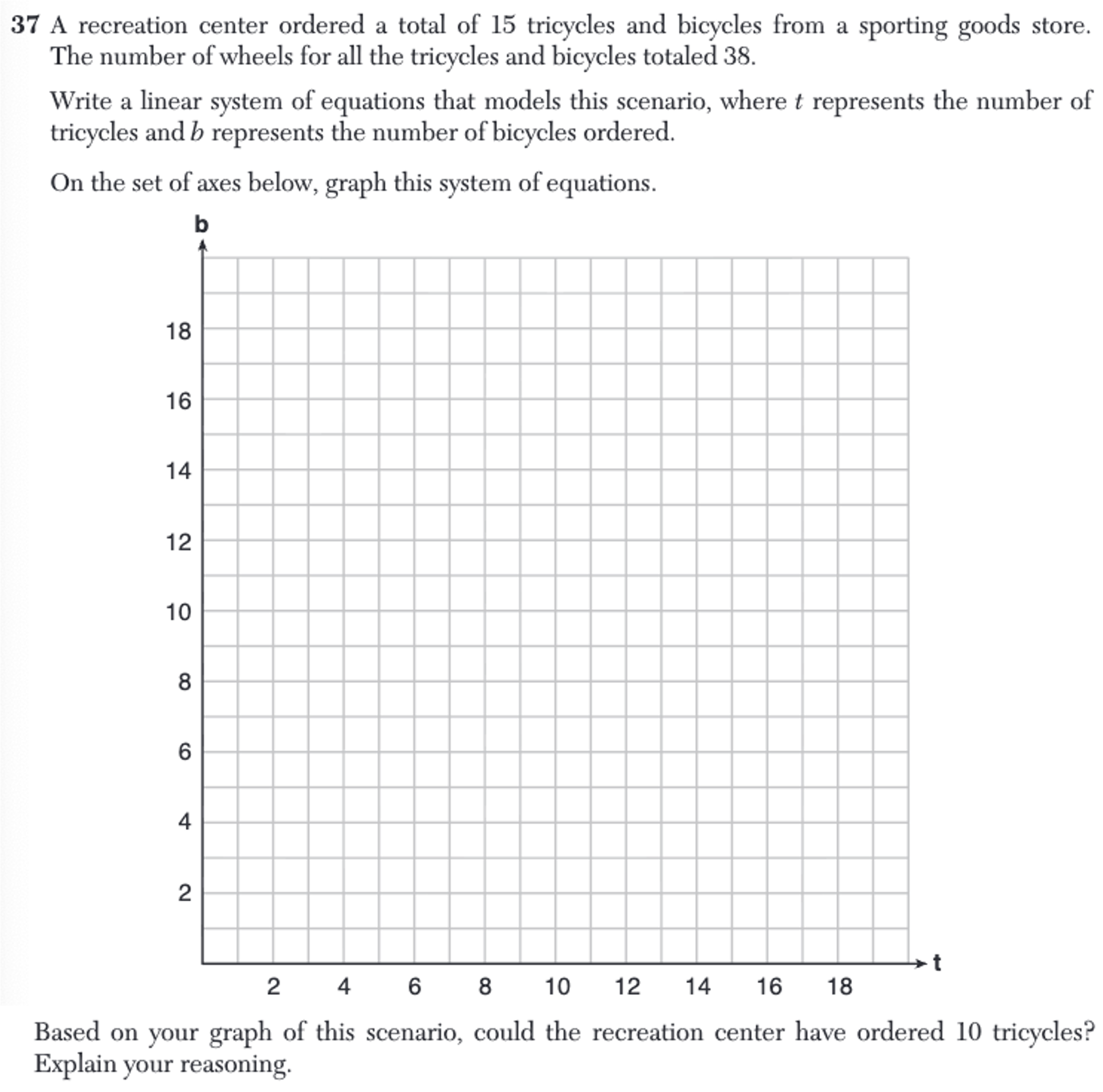}
    }
    \caption{Item (a) from the 1986 SAT math exam and Item (b) from the 2019 Regents Algebra 1 exam.  
    {\small SOURCE: \url{https://files.eric.ed.gov/fulltext/ED311087.pdf};
    \url{https://www.nysedregents.org/algebraone/119/algone12019-exam.pdf}}}\label{fig:items}
\end{figure}

Despite sustained efforts to develop fair assessments, standardized tests have long included items later flagged as biased against particular groups. Figure \ref{fig:items} presents two examples of items that may suffer from construct-irrelevant bias: one from the SAT math exam and the other from the New York State Regents Algebra I exam. The SAT is a standardized test widely used for college admissions in the United States. In contrast, the Regents examinations are statewide standardized assessment administered by the New York State Education Department to evaluate students' academic proficiency and determine graduation eligibility.

According to \Textcite{rosser1989sat}, the basketball question shown in Figure \ref{fig:SAT} exhibited the largest performance difference between males and females on the 1987 SAT; approximately 27\% more males answered it correctly than females on a marginal scale. It highlights a broader concern in the test fairness literature \parencite[e.g.,][]{rosser1989sat, zumbo2007three, chubbuck2016whos}. That is, items using stereotypically masculine contexts (e.g., sports, war, certain science or technical scenarios) tend to favor males, while items focusing on relationships or humanities frequently favor females. This basketball problem serves as a key piece of evidence for that claim. 

More recently, an item from the 2019 Regents exam in Figure \ref{fig:regent} raised fairness concerns regarding the terminology \emph{bicycles} and \emph{tricycles}. To solve this item, students must have external knowledge that a bicycle has two wheels and a tricycle has three wheels. Although it is obvious to native English speakers, ELL students, who have limited exposure to English, may fail to answer this item correctly due to vocabulary limitations rather than a lack of math ability.\footnote{\url{https://www.star-revue.com/new-york-teachers-question-regents-exams/}} Ideally, a fair assessment should allow ELL students to resolve the problem without requiring knowledge of the English prefixes \textit{bi} and \textit{tri} because linguistic proficiency is construct-irrelevant to the math ability being measured \parencite{suk_lyu_2026}.

These two motivating items illustrate how performance may be influenced by construct-irrelevant factors associated with gender and ELL status. However, when group membership (i.e., gender or ELL status) is treated as the treatment to investigate unfairness, it is non-manipulable. Therefore, identifying and interpreting the specific causes of unfairness is challenging under traditional frameworks. Throughout this paper, we use these basketball and bicycle items as motivating examples to illustrate how our separable DIF framework helps identify the underlying sources of differential functioning even when the treatment itself is non-manipulable. 

The remainder of the paper is organized as follows. We first review relevant literature on DIF, treatment manipulability, and separable effects. Next, we introduce our separable DIF framework using separable effects and differentiate between scenarios with the absence or presence of impact; we provide formal definitions and identification strategies of each separable DIF effect. We then apply the framework to the basketball item from the SAT and the bicycle item from the Regents exam. Subsequently, we introduce detection methods using causal forests and BART, and demonstrate their performance via a simulation study and a real-world application. Finally, we discuss our findings and their implications for promoting interventionist approaches in testing practices.

\section{Review}

\subsection{Test (Un)fairness: DIF}

The evaluation of test fairness has evolved since the civil rights era of the 1960s. In the bible of testing, \emph{The Standards for Educational and Psychological Testing} \parencite{aera2014standards}, fairness in testing is interpreted as ``responsiveness to individual characteristics and testing contexts so that test scores will yield valid interpretations for intended uses'' and is fundamentally viewed as a lack of measurement bias.

One of the most widely-used concepts for evaluating test fairness is DIF, which serves as a standard statistical tool for detecting items that may function unfairly against specific subgroups \parencite{holland1993differential, zumbo2007three, suk2023psychometric}. An item exhibits DIF if test-takers of identical ability, but belonging to different demographic or behavioral groups, differ in their probability of answering the item correctly \parencite{pine1977, magis2010general}. Formally, DIF is defined as
\[
P(Y_{i}=1\mid\Theta_i = \theta, A_i=a) \neq P(Y_{i}=1\mid \Theta_i = \theta, A_i=a')
\]
for a given ability level $\theta$ and distinct group values $a$ and $a'$ (e.g., focal vs. reference groups). Here, $Y_{i}$ denotes the item response of test-taker $i$ (1 for correct, 0 for incorrect), and $\Theta_{i}$ represents the test-taker's ability. The variable $A_{i}$ denotes a group or protected variable for which fairness is a concern (e.g., gender, race), and typically distinguishes between a focal group ($A_{i}=a$; e.g., females) and a reference group ($A_{i}=a'$; e.g., males). Evaluating an item for DIF requires a test-taker's ability to be known or estimated from DIF-free items, commonly referred to as anchor items. A DIF flag indicates potential bias \parencite{holland1993differential}, so items exhibiting high levels of DIF are typically revised by test developers or removed from the item bank. However, the DIF flag captures only the conditional association between group membership and item performance, rather than a causal link. 

Although gender (or other non-manipulable variables) serves as the grouping variable in the DIF analysis, psychometricians do not simply interpret DIF as an inherent effect of gender itself. Instead, they try to link gender-related DIF to specific item characteristics, such as item format, context, or content \parencite{zumbo2007three}. For example, consider the basketball item in Figure \ref{fig:SAT}. The observed DIF effect is likely not due to gender per se, but rather because the item's content (basketball) may be more engaging or familiar to boys than to girls \parencite{rosser1989sat}. Thus, researchers look beyond the uninterpretable ``gender'' effect to identify substantive causes, such as gender-specific engagement or cultural experience, as the true sources of differential functioning.

\subsection{Treatment Manipulability}

In the statistics and causal inference literature, the emphasis on manipulability for defining causal effects has a long history. This perspective aligns with \citeauthor{holland1986}'s famous (\citeyear{holland1986}) maxim, ``no causation without manipulation'' and has roots in an influential text on experimental design. \textcite{cook1979quasi}, for example, state: ``The paradigmatic assertion in causal relationships is that manipulation of a cause will result in the manipulation of an effect. \dots{} Causation implies that by varying one factor I can make another vary''  (p. 36). This school emphasizes that causal effects are well-defined only for variables that can, at least in principle, be manipulated. Thus, this principle excludes immutable characteristics, such as race or gender, as valid treatments because they cannot be meaningfully assigned to an individual. Proponents of this view are cautious about drawing causal conclusions regarding non-manipulable variables and prefer to ground causal claims in experimental interventions. 

To address the limitations of non-manipulable treatments, several researchers have proposed alternative frameworks. \Textcite{greiner2011causal} shifted the focus from the immutable characteristic itself to the \emph{perception} of that characteristic (e.g., using racialized names on resumes to manipulate the perception of race). However, this approach faces criticism that the perception of race does not capture the inherent, multidimensional nature of race itself. \Textcite{vanderweele2014causal} also argued that because race itself is not directly intervened on, causal questions should be framed in terms of interventions on more manipulable factors (e.g., socioeconomic status, education) and the pathways through which racial classifications influence outcomes. This logic aligns with the view that race is not a monolithic treatment, but rather a composite of factors. Similarly, \Textcite{sen2016race} describe race as a ``bundle of sticks'' (i.e., manifest variables) and leverage social constructivism\footnote{Social constructivism is a theory that emphasizes how knowledge, meanings, and realities are constructed through social interaction and cultural contexts, rather than being simply discovered or inherent.} to conceptualize manipulability.  Building on this view, \Textcite{weinberger2023signal} proposes a signal-based manipulation, where signals are variables that reliably track a non-manipulable treatment.

In contrast, another school of thought permits a broader conceptualization of causality, rooted in theoretical interventions \parencite{pearl2000causality, woodward2003making}. Following Judea Pearl, this school challenges the strict necessity of physical manipulability. Pearl's ``interventionist'' framework formalizes causation using the $do$-operator \parencite{pearl2000causality}, which does not strictly require human actions. These interventions can be exogenous processes that conceptually fix the value of a variable in a structural model, like the causal effect of the moon's gravitational attraction on the tides \parencite{pearl2000causality}. This perspective allows for causal claims about non-manipulable variables (e.g., moon, gender), provided they are cast as hypothetical interventions within a well-specified structural model. In this paper, we align with the former school rather than the latter to prioritize actionable interventions by human agents in identifying and rectifying the causes of test unfairness. 

\subsection{Separable Effects}\label{sec:SeparableEffects}

\textcite{robins2010alternative} introduced the concept of treatment decomposition, or separable effects, as an alternative framework for interpreting mediated effects via manipulable parameters. This framework has been further developed in subsequent work \parencite[e.g.,][]{robins2022interventionist, stensrud2021generalized, shpitser2017modeling}. Treatment decomposition splits a treatment into distinct, conceptually meaningful components, each of which acts on the outcome through different causal pathways. For example, a treatment $A$ (e.g., cigarette) can be decomposed into a drug's active ingredient (e.g., nicotine, denoted as $A_Y$) and auxiliary ingredients (e.g., non-nicotine, denoted as $A_D$), which operate on the outcome via distinct mechanisms.

\begin{figure}[!htbp]
    \centering
\subcaptionbox{DAG for treatment decomposition \label{fig:extDAG}}[.45\textwidth]{
\begin{tikzpicture}
\tikzset{line width=1pt, outer sep=1pt,
ell/.style={draw,fill=white, inner sep=3pt,
line width=1pt},
swig vsplit={gap=5pt,
inner line width right=0.5pt}};
\node[name=A,ell,shape=ellipse] at (0,0){$A$}; 
\node[name=A_D,ell,shape=ellipse] at (2,1){$A_D$};
\node[name=A_Y,ell,shape=ellipse] at (2,0){$A_Y$};
\node[name=X,ell,shape=ellipse] at (4,2){$D$};
\node[name=Y,ell,shape=ellipse] at (6, 0){$Y$};
\draw [->, line width=1.75pt] (A) to[in=-160] (A_D);
\draw [->, line width=1.75pt] (A) to (A_Y);
\draw [->, line width=0.75pt] (X) to (Y);
\draw [->, line width=0.75pt] (A_D) to (X);
\draw [->, line width=0.75pt] (A_Y) to (Y);
\end{tikzpicture}}
\hspace{0.1in}
    \subcaptionbox{SWIG for treatment decomposition \label{fig:extSWIG}}[.45\textwidth]{
\begin{tikzpicture}
\tikzset{line width=1pt, outer sep=1pt,
ell/.style={draw,fill=white, inner sep=3pt,
line width=1pt},
swig vsplit={gap=5pt,
inner line width right=0.5pt}};
\node[name=A,ell,shape=ellipse] at (0,0){$A$}; 
\node[name=A_D,shape=swig vsplit] at (2,1){  \nodepart{left}{$A_D$} \nodepart{right}{$a_D$} };
\node[name=A_Y,shape=swig vsplit] at (2,0){  \nodepart{left}{$A_Y$} \nodepart{right}{$a_Y$} };
\node[name=X,ell,shape=ellipse] at (4,2){$D^{a_D}$};
\node[name=Y,ell,shape=ellipse] at (6, 0){$Y^{a_Y, a_D}$};
\draw [->, line width=1.75pt] (A) to[in=-160] (A_D);
\draw [->, line width=1.75pt] (A) to (A_Y);
\draw [->, line width=0.75pt] (X) to (Y);
\draw [->, line width=0.75pt] (A_D) to (X);
\draw [->, line width=0.75pt] (A_Y) to (Y);
\end{tikzpicture}
    }
    \caption{Directed acyclic graph (DAG) and single-world intervention graph (SWIG) for treatment decomposition}\label{fig:extDAGSWIG}
\end{figure}

Figure \ref{fig:extDAGSWIG} illustrates this treatment decomposition using a directed acyclic graph \parencite[DAG;][]{pearl2000causality} and a single-world intervention graph \parencite[SWIG;][]{richardson2013swig}\footnote{SWIG is a causal framework that unifies causal graphs and potential outcomes and is based on the Finest Fully Randomized Causally Interpreted Structural Tree Graph (FFRCISTG) model \parencite{robins1986ffrcistg, richardson2013swig}.}. In these graphs, the bold arrows from $A$ to $A_Y$ and $A_D$ indicate a deterministic relationship, i.e.,  $A \equiv A_Y \equiv A_D$. Figure \ref{fig:extDAG} represents the causal structure in an \emph{extended DAG} \parencite{robins2010alternative}, which is a transformation of the original causal DAG. It includes the treatment $A \in \{0, 1\}$ and its components $A_Y  \in \{0, 1\}$ and $A_D  \in \{0, 1\}$, encoding the mechanisms by which these components individually operate on the outcomes. Figure \ref{fig:extSWIG} presents the corresponding SWIG, where each component node ($A_Y$ or $A_D$) is split into random and fixed halves. In this single world, all individuals were hypothetically assigned to the component values $A_Y=a_Y$ and $A_D=a_D$; thus, the potential mediator and potential outcome of interest become $D^{a_D}$ and $Y^{a_Y, a_D}$, respectively. 

Based on Figure \ref{fig:extSWIG}, separable effects are defined as causal effects attributable to distinct treatment components. First, the \emph{separable direct effect} (SDE) is defined as:
\begin{align*}
\text{SDE}(a_D) = E[Y^{1,a_D} - Y^{0,a_D}].
\end{align*}
This estimand measures the expected difference in the outcome when changing $A_Y$ from 0 to 1, while holding $A_D$ constant at level $a_D$. For the cigarette example, it represents the effect of the nicotine component on myocardial infarction (outcome) while the non-nicotine component is held fixed. Second, the \emph{separable indirect effect} (SIE) is defined as:
\begin{align*}
\text{SIE}(a_Y) = E[Y^{a_Y,1} - Y^{a_Y,0}].
\end{align*}
This measures the expected difference in the outcome when changing $A_D$ from 0 to 1, holding $A_Y$ constant at level $a_Y$. Continuing the cigarette example, it indicates the effect of the non-nicotine component while holding the nicotine component fixed. 

Additionally, \textit{conditional separable effects} can be defined \parencite{Stensrud2023CSE}. For example, the conditional SDE (C-SDE) is written with $D=d \in \{0, 1\}$ as:
\begin{align}\label{eq:C-SDE}
\text{C-SDE}(a_D, d) = E[Y^{1,a_D} - Y^{0,a_D} \mid D^{a_D} = d].
\end{align}
This measures the SDE conditional on the mediator under the intervention $A_D=a_D$. In the cigarette example, it represents the effect of the nicotine component conditional on hypertension (the mediator) status while holding the non-nicotine component fixed. 

These separable effects differ from natural (or pure) direct and indirect effects, and controlled direct effects \parencite{robins1992identifiability, pearl2001direct}. Those traditional estimands typically require well-defined treatments and mediators, as well as cross-world counterfactuals (e.g., $Y^{A=1, D^{A=0}}$). In practice, direct interventions on mediators (required for controlled effects) are often infeasible, and cross-world interventions (required for natural effects) are not experimentally implementable. In contrast, the separable effect formulation requires only that the interventions on $A_D$ and $A_Y$ be substantively meaningful. This allows for straightforward discussions with subject matter experts (e.g., item developers) about feasible experimental interventions \parencite{robins2022interventionist}. 

Separable effects have been extended to various settings, including spillover effects \parencite{shpitser2017modeling}, competing events \parencite{stensrud2021generalized, stensrud2022separable}, and conditional effects with post-treatment events \parencite{Stensrud2023CSE}. However, there has been limited discussion on their potential for handling non-manipulable treatments. In this paper, we utilize a type of conditional separable effect to define meaningful estimands in the DIF context. See the next section for details.  

\section{Separable DIF with Intervening Variables}

\subsection{Simple Separable DIF}

In this subsection, we assume that there is no effect of the non-manipulable treatment $A$ on ability $\Theta$ (i.e., no impact). Confounders $X$ are permitted to influence both $A$ and $Y$ when treatment $A$ is endogenous (e.g., ELL status), whereas $X = \varnothing$ if $A$ is exogenous (e.g., gender). We provide the definition of this separable DIF estimand and identification strategies for the defined effect.

\subsubsection{Definition}

Figure \ref{fig:DAG_A} illustrates a SWIG on a non-manipulable treatment $A$ (e.g., gender, race, ELL status), where the blue arrow indicates an unfair, nonzero effect. In Supplemental Appendix \ref{app:DAG}, we provide the corresponding DAGs for each SWIG used throughout this section and the subsequent sections. Figure \ref{fig:DAG_A} implies the question, ``What would the item response have been had your gender been different?'' As discussed, this question is  conceptually ambiguous and experimentally infeasible.

\begin{figure}[!htbp]
\centering
    \subcaptionbox{SWIG with $A$  \label{fig:DAG_A}}{
\begin{tikzpicture}
\tikzset{line width=1pt, outer sep=1pt,
ell/.style={draw,fill=white, inner sep=3pt,
line width=1pt},
swig vsplit={gap=5pt,
inner line width right=0.5pt},
ellempty/.style={fill=white}};
\node[name=A,shape=swig vsplit] at (0,0){  \nodepart{left}{$A$} \nodepart{right}{$a$} }; 
\node[name=Theta,ell,shape=ellipse] at (2,2){$\Theta$};
\node[name=Y,ell,shape=ellipse] at (4, 0){$Y^a$};
\node[name=X,ell,  shape=ellipse] at (2,-1.3) {$X$};

\draw [->, line width=0.75pt] (Theta) to (Y);
\draw [draw=blue, ->, line width=0.75pt] (A) to (Y);

\draw [->, line width=0.75pt] (X) to[in=-105, out=190] (A);
\draw [->, line width=0.75pt] (X) to[in=-100, out=-10] (Y);
\end{tikzpicture}}
\hspace{0.5in}
    \subcaptionbox{SWIG with $A$ and $A_Y$ \label{fig:DAG_AY}}{
\begin{tikzpicture}
\tikzset{line width=1pt, outer sep=1pt,
ell/.style={draw,fill=white, inner sep=3pt,
line width=1pt},
swig vsplit={gap=5pt,
inner line width right=0.5pt},
ellempty/.style={fill=white}};
\node[name=A,ell,shape=ellipse] at (-0.1,0){$A$}; 
\node[name=A_Y,shape=swig vsplit] at (2,0){\nodepart{left}{$A_Y$} \nodepart{right}{$a_Y$} };
\node[name=Theta,ell,shape=ellipse] at (3,2){$\Theta$};
\node[name=Y,ell,shape=ellipse] at (5, 0){$Y^{a_Y}$};
\node[name=X,ell,  shape=ellipse] at (2.5,-1.3) {$X$};

\draw [->, line width=1.75pt] (A) to (A_Y);
\draw [->, line width=0.75pt] (Theta) to (Y);
\draw [draw=blue, ->, line width=0.75pt] (A_Y) to (Y);

\draw [->, line width=0.75pt] (X) to[in=-80, out=190] (A);
\draw [->, line width=0.75pt] (X) to[in=-100, out=-10] (Y);
\end{tikzpicture}}
\caption{Single-world intervention graphs (SWIGs) with non-manipulable treatment $A$, ability $\Theta$, potential item response $Y^{(\cdot)}$, confounders $X$, and manipulable treatment component $A_Y$. \small{The blue arrow indicates an unfair path, and the bold arrow indicates a deterministic relationship, $A\equiv A_Y$.}}\label{fig:DAG_no_impact}
\end{figure}

Instead, Figure \ref{fig:DAG_AY} illustrates an extended SWIG that explicitly models an intervening variable $A_Y$ of the non-manipulable treatment $A$. In the observed world, there is a deterministic relationship, $A \equiv A_Y$, indicated by the bold arrow. 
In the DIF context, we define $A_Y$ as follows:
\begin{definition}\label{def:A_Y}
$A_Y$ is a construct-irrelevant component of $A$ that affects the response $Y$ (e.g., experience, skill).    
\end{definition}
\noindent That is, $A_Y$ is a manipulable component that reliably tracks status $A$ while directly influencing the item response.\footnote{In the signal-based manipulation proposed by \textcite{weinberger2023signal}, a \textit{signal} is a manipulable variable that reliably tracks an underlying non-manipulable treatment, and interventions on the signal are used to learn about the effect of that underlying treatment itself. However, our approach uses a treatment component and targets the separable effect of a given component, rather than the total effect of the non-manipulable treatment.} By shifting attention from $A$ to $A_Y$, we can manipulate the intervening variable $A_Y$, and ask a substantively meaningful question, ``What would the item response have been if the gender-linked contextual component were different?''

For example, while it is impossible to manipulate gender itself, we can use a construct-irrelevant component $A_Y$ that has a deterministic relationship with $A$. For example, in the basketball example, being a boy implies high societal exposure to male-dominated sports, whereas being a girl implies low exposure. This deterministic relation likely held strictly in the 1980s when the test was administered. Similarly, in the bicycle item, it assumes ELL status implies low English lexical proficiency. Unlike $A$, we can design an intervention on $A_Y$; for instance, we could provide a brief tutorial on basketball rules (effectively setting $A_Y=a_Y$ for everyone) or provide a confusing or irrelevant lesson (setting $A_Y=a_Y'$). The ability to manipulate the \emph{exposure} component allows researchers to design well-defined experiments in principle. 

With a meaningful intervening variable $A_Y$, we can define a separable effect in the absence of the mediator/ability pathway, which we term \textit{simple separable DIF} as follows:
\begin{definition}\label{def:SS}
(Simple Separable DIF) An item has simple separable DIF if 
\begin{center}
$P(Y^{a_Y} = 1 \mid \Theta=\theta) \neq P(Y^{a_Y'} = 1 \mid \Theta=\theta)$ \\   
\end{center}
\noindent for some ability level $\theta$ and distinct values $a_Y$ and $a_Y'$ attainable by $A_Y$.
\end{definition}
\noindent Here, $P(Y^{a_Y} = 1 \mid \Theta=\theta)$ is the counterfactual probability of endorsing an item if all examinees were assigned to the component level $A_Y=a_Y$, conditional on having the same ability $\Theta=\theta$. The separable DIF effect of the intervening variable is the interventionist estimand of actual interest.\footnote{One could also define \textit{simple separable DIF on the focal group} as: $P(Y^{A_Y=a} = 1 \mid \Theta=\theta, A_Y=a) \neq P(Y^{A_Y=a'} = 1 \mid \Theta=\theta, A_Y=a)$. This estimand focuses on the separable effect within the focal population.}

\subsubsection{Identification}

In the observed data, we impose the determinism condition: $A \equiv A_Y$.  To identify the simple separable DIF effect under such a setting, we require the following assumptions:

\begin{itemize}

\item[\HT{A1}] Consistency: If $A=a$, $Y=Y^{a_Y=a} \ $ for $a\in\{0,1\}$
        
\item[\HT{A2}] Ignorability: $ A  \indep Y^{a_Y=a} \cond X, \Theta \ $ for $a\in\{0,1\}$

\item[\HT{A3}] Positivity: $\Pr ( A=a \cond X, \Theta ) > 0 \ $ for $a \in \{0,1\}$

\item[\HT{A4}] Dismissible Component Conditions (future trial $G$): 
\begin{itemize}
  \item[(i)] $Y(G) \indep A(G) \cond A_{Y}(G), X(G), \Theta(G)$
  \item[(ii)] $\Theta(G) \indep A_{Y}(G) \cond A(G), X(G)$
\end{itemize}  
\end{itemize} 
Here, Assumption (A1) links the observed outcome to the potential outcome under the component level $A_Y=a$. This implies that there are no multiple versions of the intervening component and no interference between test-takers. Assumption (A2) states that, within strata defined by pre-treatment covariates $X$ and ability $\Theta$, the group $A$ is as-if randomly assigned, and thus independent of the potential outcome $Y^{a_Y=a}$. Assumption (A3) requires that for every combination of covariates and ability, there is a nonzero probability of observing members from each group.  

In Assumption (A4), ``$(G)$'' refers to a hypothetical future trial where $A_Y$ is randomly assigned \parencite{stensrud2021generalized}, as illustrated in Figure \ref{fig:DAG_G1}. In this trial $G$, researchers can design a well-defined intervention on the component $A_Y$ (e.g., providing a correct lesson on basketball rules versus an incorrect lesson). Due to this random assignment, $A_Y$ is no longer influenced by $A$, and thus, no arrow exists between $A$ and $A_Y$ in Figure \ref{fig:DAG_G1}. From this figure, conditions of Assumption (A4) can be read off. Specifically, condition (i) states that the non-manipulable treatment $A$ is independent of the response $Y$ conditional on $A_Y$, $X$, and $\Theta$ in the future trial $G$.\footnote{From Figure \ref{fig:DAG_G1}, $Y(G) \indep A(G) \cond A_{Y}(G), X(G)$ holds, regardless of whether $\Theta$ is conditioned on or not.} In our basketball example, this means that gender is independent of the item response conditional on gender-specific exposure, ability, and relevant confounders. Condition (ii) states that the component $A_Y$ is independent of ability $\Theta$, conditional on $A$ and $X$. In the basketball example, this means that gender-specific exposure is independent of ability conditional on gender and relevant confounders. Under the FFRCISTG model, condition (ii) ensures $A_Y$ partial isolation \parencite{stensrud2021generalized}, i.e., $\Theta^{a_Y}=\Theta^{a_Y'}$, which is trivially maintained in the absence of impact. See Section ``\nameref{sec:con}'' for more discussion on this assumption.

\begin{figure}[!htbp]
    \centering
    \begin{tikzpicture}
\tikzset{line width=1pt, outer sep=1pt,
ell/.style={draw,fill=white, inner sep=3pt,
line width=1pt},
swig vsplit={gap=5pt,
inner line width right=0.5pt},
ellempty/.style={fill=white}};
\node[name=A,ell,shape=ellipse] at (-1,0){$A(G)$}; 
\node[name=A_Y,ell,shape=ellipse] at (1.5,0){$A_Y(G)$};
\node[name=Theta,ell,shape=ellipse] at (3,2){$\Theta(G)$};
\node[name=Y,ell,shape=ellipse] at (5, 0){$Y(G)$};
\node[name=X,ell,  shape=ellipse] at (2,-1.5) {$X(G)$};

\draw [->, line width=0.75pt] (Theta) to (Y);
\draw [draw=blue, ->, line width=0.75pt] (A_Y) to (Y);

\draw [->, line width=0.75pt] (X) to[in=-80, out=190] (A);
\draw [->, line width=0.75pt] (X) to[in=-100, out=-10] (Y);
\end{tikzpicture}
    \caption{Transformed directed acyclic graph from Figure \ref{fig:DAG_AY} under a future trial $G$.}\label{fig:DAG_G1}
\end{figure}

Based on the above assumptions, we can identify the simple separable DIF effect $\tau_{\rm SS}(\theta)$ as:
\begin{align}\label{eq:SS_iden}
\tau_{\rm SS}(\theta) & \coloneqq E(Y^{a_Y=a} \mid \Theta=\theta)
- E(Y^{a_Y=a'} \mid \Theta=\theta) \nonumber \\ 
&= E\{ E( Y \mid  A=a, \Theta=\theta, X) \mid \Theta=\theta \}
- E\{ E( Y \mid  A=a', \Theta=\theta, X) \mid \Theta=\theta \}. 
\end{align}
This result is equivalent to the standard $g$-formula (adjustment formula) for the conditional average treatment effect (CATE) of $A$ on $Y$ within ability level $\Theta=\theta$. See Supplemental Appendix \ref{app:iden} for the detailed proof. Under Assumptions (A1)--(A4), we interpret Equation \eqref{eq:SS_iden} as the effect of the manipulable intervening component $A_Y$. 

\raggedbottom

\subsection{General Separable DIF}

In this subsection, we assume the presence of impact \parencite{ackerman1992didactic}, where the non-manipulable treatment $A$ directly affects the ability $\Theta$ (i.e., $A \rightarrow \Theta$). In this setting, non-manipulable treatment $A$ (e.g., gender, race) affects both ability $\Theta$ and item response $Y$, and $\Theta$ serves as a mediator. In testing practices, only the direct path from $A$ to $Y$ is considered unfair, while the indirect path $A \rightarrow \Theta \rightarrow Y$ is permissible. That is, only the direct path of $A$ on $Y$ refers to a DIF effect, representing a nonzero effect of $A$ in an assumed outcome model for $Y$. We provide the definition of the separable DIF estimand for this setting and discuss identification strategies.

\subsubsection{Definition}\label{sec:td}

Figure \ref{fig:SWIG_A_Impact} illustrates a SWIG where interventions (represented by node-splitting) occur in both non-manipulable treatment $A$ and ability $\Theta$, and the covariates $W$ contain all common causes between $A$, $\Theta$, and $Y$. As mentioned earlier, intervening on a non-manipulable treatment or on ability scores is infeasible in practice. Instead, Figure \ref{fig:SWIG_AY_AD_Impact} decomposes the non-manipulable treatment $A$ into manipulable components $A_Y$ and $A_\Theta$, which operate on the outcome along distinct causal pathways. In this framework, we intervene on $A_Y$ and $A_\Theta$ instead of $A$ and $\Theta$.

\begin{figure}[!htbp]
\centering
    \subcaptionbox{SWIG with $A$ \label{fig:SWIG_A_Impact}}{
\begin{tikzpicture}
\tikzset{line width=1pt, outer sep=1pt,
ell/.style={draw,fill=white, inner sep=3pt,
line width=1pt},
swig vsplit={gap=5pt,
inner line width right=0.5pt},
ellempty/.style={fill=white}};
\node[name=A,shape=swig vsplit] at (0,0){\nodepart{left}{$A$} \nodepart{right}{$a$} }; 
\node[name=Theta,shape=swig vsplit] at (2,1.6){\nodepart{left}{$\Theta^a$} \nodepart{right}{$\theta$} };
\node[name=Y,ell,shape=ellipse] at (4, 0){$Y^{a,\theta}$};
\node[name=X,ell,  shape=ellipse] at (1.5,2.8) {$W$};

\draw [->, line width=0.75pt] (Theta) to (Y);
\draw [draw=blue, ->, line width=0.75pt] (A) to (Y);

\draw [->, line width=0.75pt] (X) to[in=120, out=200] (A);
\draw [->, line width=0.75pt] (X) to[in=100, out=300] (Theta);
\draw [->, line width=0.75pt] (A) to (Theta);
\draw [->, line width=0.75pt] (X) to[in=80, out=-10] (Y);
\end{tikzpicture}}
\hspace{0.5in}
    \subcaptionbox{SWIG with $A$, $A_Y$, and $A_\Theta$ \label{fig:SWIG_AY_AD_Impact}}{
\begin{tikzpicture}
\tikzset{line width=1pt, outer sep=1pt,
ell/.style={draw,fill=white, inner sep=3pt,
line width=1pt},
swig vsplit={gap=5pt,
inner line width right=0.5pt},
ellempty/.style={fill=white}};
\node[name=A,ell,shape=ellipse] at (-0.5,0){$A$}; 
\node[name=A_Y,shape=swig vsplit] at (2,0){\nodepart{left}{$A_Y$} \nodepart{right}{$a_Y$} };
\node[name=A_D,shape=swig vsplit] at (1.5,0.8){\nodepart{left}{$A_\Theta$} \nodepart{right}{$a_\Theta$} };
\node[name=Theta,ell,shape=ellipse] at (3.5,1.6){$\Theta^{a_\Theta}$};
\node[name=Y,ell,shape=ellipse] at (5.5, 0){$Y^{a_Y, a_\Theta}$};
\node[name=X,ell,  shape=ellipse] at (3,2.8) {$W$};

\draw [->, line width=1.75pt] (A) to (A_Y);
\draw [->, line width=1.75pt] (A) to (A_D);
\draw [->, line width=0.75pt] (Theta) to (Y);
\draw [draw=blue, ->, line width=0.75pt] (A_Y) to (Y);

\draw [->, line width=0.75pt] (X) to[in=100, out=190] (A);
\draw [->, line width=0.75pt] (X) to[in=100, out=300] (Theta);
\draw [->, line width=0.75pt] (A_D) to (Theta);
\draw [->, line width=0.75pt] (X) to[in=60, out=-10] (Y);
\end{tikzpicture}}
\caption{Single-world intervention graphs (SWIGs) with non-manipulable treatment $A$, potential ability $\Theta^{(\cdot)}$, potential item response $Y^{(\cdot)}$, covariates $W$, and manipulable treatment components $A_Y$ and $A_\Theta$. \small{The blue arrow indicates an unfair path, and the bold arrow indicates a deterministic relationship, $A \equiv A_Y \equiv A_\Theta$.}}\label{fig:SWIG_impact}
\end{figure}
While the definition of $A_Y$ (i.e., Definition \ref{def:A_Y}) remains the same, we define the second component $A_\Theta$ as:
\begin{definition}
$A_\Theta$ is a construct-relevant component of $A$ that does not affect the response $Y$, except through ability $\Theta$.    
\end{definition}
\noindent Specifically, $A_\Theta$ is a manipulable component of $A$ that reliably tracks test-takers' status $A$ and affects $Y$ only via $\Theta$. In the observed world, there is a deterministic relationship, $A \equiv A_Y \equiv A_\Theta$, as highlighted by the bold arrows in Figure \ref{fig:SWIG_AY_AD_Impact}.

As a concrete example, consider the bicycle item from Figure \ref{fig:regent} in our motivating examples. Here, $A$ represents ELL status. $A_Y$ represents English lexical proficiency, while $A_\Theta$ represents instructional English comprehension. Both components reliably track $A$. Unlike $A$, we can intervene on $A_Y$ and $A_\Theta$. For instance, an intervention on $A_Y$ can be a brief pre-test session on the specific vocabulary prefixes (e.g., linking `bi-' to 2 and `tri-' to 3). Similarly, an intervention on $A_\Theta$ can be to assign a teacher who instructs mathematics in the student's primary language versus one who instructs in their second language. In this framework, $A_\Theta$ and $A_Y$ are well-defined, manipulable variables, even if the status $A$ itself is not.

Using these two intervening variables, we define a separable effect in the presence of impact, which we term \textit{general separable DIF}. Let $Y^{a_Y,a_\Theta}$ be the potential outcome if a test-taker were assigned to component levels $A_Y=a_Y$ and $A_\Theta=a_\Theta$, and let $\Theta^{a_\Theta}$ be the potential ability if they were assigned to the level $A_\Theta=a_\Theta$. 
\begin{definition}\label{def:GS}
(General Separable DIF) An item has general separable DIF if 
\begin{align*}
P(Y^{a_Y, a_\Theta}=1 \mid \Theta^{ a_\Theta}=\theta ) \neq P(Y^{a_Y', a_\Theta}=1 \mid \Theta^{a_\Theta}=\theta ),
\end{align*}
\noindent for some ability level $\theta$, distinct values $a_Y$ and $a_Y'$ attainable by $A_Y$, and value $a_\Theta$ attainable by $A_\Theta$.
\end{definition}
\noindent Here, $P(Y^{a_Y, a_\Theta} = 1 \mid \Theta^{a_\Theta}=\theta)$ represents the counterfactual probability of endorsing an item if all test-takers were assigned to the components $A_Y=a_Y$ and $A_\Theta=a_\Theta$, conditional on their potential ability under the same assignment. This definition aligns with the conditional separable effects (i.e., Equation \eqref{eq:C-SDE}) in the literature.

\subsubsection{Identification}

In the observed data, we impose the determinism condition: $A \equiv A_Y \equiv A_\Theta$. In this setting, we require the following assumptions to identify the general separable DIF effect.
\begin{itemize}
        
    \item[\HT{B1}] Consistency: If $A=a$, then $Y = Y^{a_Y=a, a_\Theta=a} \ \text{and} \ \Theta = \Theta^{a_\Theta=a} \ $ for $a \in \{0, 1\}$

    \item[\HT{B2}] Ignorability: $(Y^{a_Y=a, a_\Theta=a}, \Theta^{a_\Theta=a})\indep A \cond W \ $ for $a\in\{0,1\}$

    \item[\HT{B3}] Positivity: $f( \Theta=\theta \cond A=a,W )>0$ and $\Pr ( A=a \cond W ) > 0 \ $ for $a \in \{0,1\}$

    \item[\HT{B4}] Dismissible Component Conditions (future trial $G$): 
    \begin{itemize}
        \item[(i)] $Y(G) \indep A_{\Theta}(G) \cond A_{Y}(G), \Theta(G), W(G)$ 
        \item[(ii)] $\Theta(G) \indep A_{Y}(G) \cond A_{\Theta}(G), W(G)$
    \end{itemize}    
\end{itemize}
Specifically, Assumption (B1) states that the observed outcome and ability are linked to the corresponding potential outcome and potential mediator under $A_Y=A_\Theta=a$, respectively. Assumption (B2) states that, within every value of the covariates $W$, the treatment variable $A$ is as-if randomly assigned and thus independent of the potential outcomes $Y^{a_Y=a, a_\Theta=a}$ and the potential mediator $\Theta^{a_\Theta=a}$. Assumption (B3) states that, for every value of the covariates $W$, there is nonzero density for observing $\Theta=\theta$ within each treatment group, and there is nonzero probability of observing $A=0$ or $A=1$ given $W$.

In Assumption (B4), ``$(G)$'' refers to a future trial where the components $A_Y$ and $A_\Theta$ are intervened on separably \parencite{stensrud2021generalized}. In this trial $G$, researchers can randomly assign $A_Y$ (e.g., providing a targeted vocabulary lesson versus a control) and $A_\Theta$ (e.g., providing a math class in the primary language versus a second language). We illustrate this structure in Figure \ref{fig:DAG_G2}, where the conditions of Assumption (B4) can be read off. Condition (i) states that the component $A_\Theta$ is independent of the response $Y$ conditional on $A_Y$, $\Theta$, and $W$. In our bicycle example, this condition means that instructional English comprehension is independent of $Y$ conditional on English lexical proficiency, algebra ability, and relevant confounders. Condition (ii) states that the component $A_Y$ is independent of ability $\Theta$ conditional on $A_\Theta$ and $W$. In the bicycle example, this means that English lexical proficiency is independent of algebra ability conditional on instructional English comprehension and relevant confounders. Under the FFRCISTG model, these conditions encode the partial isolation of the two components \parencite{stensrud2021generalized}, e.g., $A_Y$ partial isolation ($\Theta^{a_Y, a_\Theta}=\Theta^{a_Y', a_\Theta}$) and $A_\Theta$ partial isolation ($Y^{a_Y, a_\Theta}=Y^{a_Y, a_\Theta'}$ if $\Theta^{a_\Theta} = \Theta^{a_\Theta'}$). See Section ``\nameref{sec:con}'' for further discussion on this assumption.

\begin{figure}[!htbp]
    \centering
    \begin{tikzpicture}
\tikzset{line width=1pt, outer sep=1pt,
ell/.style={draw,fill=white, inner sep=3pt,
line width=1pt},
swig vsplit={gap=5pt,
inner line width right=0.5pt},
ellempty/.style={fill=white}};
\node[name=A,ell,shape=ellipse] at (-1,0){$A(G)$}; 
\node[name=A_Y,ell,shape=ellipse] at (2,0){$A_Y(G)$};
\node[name=A_D,ell,shape=ellipse] at (.8,1){$A_\Theta(G)$};
\node[name=Theta,ell,shape=ellipse] at (3,1.8){$\Theta(G)$};
\node[name=Y,ell,shape=ellipse] at (6, 0){$Y(G)$};
\node[name=X,ell,  shape=ellipse] at (2.5,3.2) {$W(G)$};

\draw [->, line width=0.75pt] (Theta) to (Y);
\draw [->, line width=0.75pt] (A_D) to (Theta);
\draw [draw=blue, ->, line width=0.75pt] (A_Y) to (Y);

\draw [->, line width=0.75pt] (X) to[in=80, out=190] (A);
\draw [->, line width=0.75pt] (X) to[in=100, out=-10] (Y);
\draw [->, line width=0.75pt] (X) to[in=100, out=300] (Theta);
\end{tikzpicture}
    \caption{Transformed directed acyclic graph from Figure \ref{fig:SWIG_AY_AD_Impact} under a future trial $G$.}\label{fig:DAG_G2}
\end{figure}

Based on the above assumptions, we can identify the general separable DIF effect $\tau_{\rm GS}(\theta)$ as:
\begin{align}
\tau_{\rm GS}(\theta) &\coloneq E(Y^{a_Y=a, a_\Theta=a} \mid \Theta^{ a_\Theta=a}=\theta ) - E(Y^{a_Y=a', a_\Theta=a} \mid \Theta^{ a_\Theta=a}=\theta ) \nonumber \\
 &= \frac{E_{W} \big[\{E(Y \mid A = a , \Theta = \theta, W) - E(Y \mid A = a', \Theta = \theta, W)\} f(\Theta = \theta \mid A = a, W)\big]}{E_{W} [f(\Theta = \theta \mid A = a, W)]}. \label{eq:iden_GS}
\end{align}
Under Assumptions (B1)--(B4), we interpret Equation \eqref{eq:iden_GS} as the effect of changing the construct-irrelevant component $A_Y$ while holding the construct-relevant component fixed. See Supplemental Appendix \ref{app:iden} for the detailed proof.

\section{Applying Separable DIF to the Motivating Examples}

This section demonstrates the application of our separable DIF framework to the motivating examples introduced in Section ``\nameref{sec:mot_ex}.'' With our domain knowledge, we investigate the presence of simple separable DIF in the basketball item and examine the presence of general separable DIF in the bicycle item.

\subsection{Basketball Item}

The basketball item from Figure \ref{fig:SAT} intends to measure students' ability in algebra. While certain mathematical domains (e.g., spatial geometry) have historically exhibited gender differences, standard ability in algebra is generally comparable across gender groups \parencite{hyde2008gender}. That is, the distribution of the target ability does not differ between the gender groups. As a result, we assume the absence of impact and test for the presence of simple separable DIF.

Figure \ref{fig:DAG_basketball} visualizes the assumed SWIG for the basketball item, which includes student gender $A$, gender-specific socialization/exposure $A_Y$ (specifically, exposure to male-dominated sports), algebra ability $\Theta$, and potential item response $Y^{a_Y}$. Since gender is generally an exogenous variable, it is reasonable to assume that there are no confounders affecting gender $A$ and potential item response $Y^{a_Y}$. Given the absence of impact, we assume that $A_Y$ does not influence $\Theta$. We also assume that there are no construct-irrelevant components of $A$ other than gender-specific exposure, and no additional variables that influence $Y^{a_Y}$\footnote{For this item, it is difficult to imagine other construct-irrelevant variables that affect the item response beyond gender.}; thus, we effectively isolate this specific cause of differential functioning. Assuming no confounding covariates, we identify this separable DIF effect, and attribute the source of bias to gender-specific exposure.

\begin{figure}[!htbp]
\centering
\begin{tikzpicture}
\tikzset{line width=1pt, outer sep=1pt,
ell/.style={draw,fill=white, inner sep=3pt,
line width=1pt},
swig vsplit={gap=5pt,
inner line width right=0.5pt},
ellempty/.style={fill=white}};
\node[name=A,ell,shape=ellipse] at (0,0) [label = below:{\footnotesize Gender} ] {$A$}; 
\node[name=A_Y,shape=swig vsplit] at (2.6,0) [label = above:{\footnotesize {\shortstack{Gender-specific\\exposure}}} ] {\nodepart{left}{$A_Y$} \nodepart{right}{$a_Y$} };
\node[name=Theta,ell,shape=ellipse] at (5,2) [label = above:{\footnotesize Algebra ability} ] {$\Theta$};
\node[name=Y,ell,shape=ellipse] at (7, 0) [label = below:{\footnotesize Item response} ] {$Y^{a_Y}$};

\draw [->, line width=1.75pt] (A) to (A_Y);
\draw [->, line width=0.75pt] (Theta) to (Y);
\draw [draw=blue, ->, line width=0.75pt] (A_Y) to (Y);
\end{tikzpicture}
\caption{Single-world intervention graph with gender $A$, gender-specific exposure $A_Y$, algebra ability $\Theta$, and potential item response $Y^{a_Y}$. \small{The blue arrow indicates an unfair path, and the bold arrow indicates a deterministic relationship, $A\equiv A_Y$.}}\label{fig:DAG_basketball}
\end{figure}

Note that if researchers have strong theoretical reasons to suspect potential confounders between $A$ and item response $Y$, they can introduce confounders $X$ into the SWIG in Figure \ref{fig:DAG_basketball} and identify the simple separable DIF by controlling for $X$. 

\subsection{Bicycle Item}

The bicycle item from Figure \ref{fig:regent} also intends to measure students' algebra ability. However, ELL students often have structural disadvantages in learning mathematics due to their limited English comprehension, and thus exhibit a different ability distribution from non-ELL students (i.e., presence of impact). In addition, ELL students may be disadvantaged by the specific linguistic demands of the item (here, pre-fixes, `bi'- and `tri-'), which require lexical knowledge of numerical concepts that is unrelated to the target algebra ability. Given the presence of impact, we test whether this item exhibits general separable DIF.

Figure \ref{fig:DAG_bicycle} visualizes the assumed SWIG for the bicycle item. It includes a student's ELL status $A$, English lexical proficiency $A_Y$, instructional English comprehension $A_\Theta$, potential algebra ability $\Theta^{a_\Theta}$, potential item response $Y^{a_Y, a_\Theta}$, and a set of confounders $W=\{\text{SES},\text{IS}\}$. In particular, we are concerned with confounders between $\Theta^{a_\Theta}$ and $Y^{a_Y, a_\Theta}$, and those between $A$ and $Y^{a_Y, a_\Theta}$. Following prior work by \Textcite{suk_lyu_2026}, socio-economic status (SES) serves as a potential confounder affecting $\Theta^{a_\Theta}$ and $Y^{a_Y, a_\Theta}$, and immigration status (IS) serves as a potential confounder affecting $A$ and $Y^{a_Y, a_\Theta}$. Therefore, when SES and IS are measured and adjusted for, we can identify the general separable DIF effect in this item, and attribute the source of bias to English lexical proficiency. See Section \ref{sec:realdata} for a relevant real-world application.

\begin{figure}[!htbp]
\centering
\begin{tikzpicture}
\tikzset{line width=1pt, outer sep=1pt,
ell/.style={draw,fill=white, inner sep=3pt,
line width=1pt},
swig vsplit={gap=5pt,
inner line width right=0.5pt},
ellempty/.style={fill=white}};
\node[name=A,ell,shape=ellipse] at (0,0) [label = below:{\footnotesize ELL} ] {$A$}; 
\node[name=A_Y,shape=swig vsplit] at (2.65,0) [label = below:{\footnotesize {\shortstack{Lexical\\proficiency}}} ] {\nodepart{left}{$A_Y$} \nodepart{right}{$a_Y$} };
\node[name=A_D,shape=swig vsplit] at (2.35,1) [label = above:{\footnotesize {\shortstack{Instructional\\comprehension}}} ] {\nodepart{left}{$A_\Theta$} \nodepart{right}{$a_\Theta$} };
\node[name=Theta,ell,shape=ellipse] at (4.5,2) [label = below:{\footnotesize {\shortstack{Algebra\\ability}}} ] {$\Theta^{a_\Theta}$};
\node[name=Y,ell,shape=ellipse] at (7, 0) [label = below:{\footnotesize Item response} ] {$Y^{a_Y, a_\Theta}$};
\node[name=X,ell,  shape=ellipse] at (4,3.5) {IS};
\node[name=X2,ell,  shape=ellipse] at (6.5,3) {SES};

\draw [->, line width=1.75pt] (A) to (A_Y);
\draw [->, line width=1.75pt] (A) to (A_D);
\draw [->, line width=0.75pt] (Theta) to (Y);
\draw [draw=blue, ->, line width=0.75pt] (A_Y) to (Y);

\draw [->, line width=0.75pt] (X) to[in=100, out=190] (A);
\draw [->, line width=0.75pt] (X) to[in=110, out=-10] (Y);

\draw [->, line width=0.75pt] (X2) to[in=60, out=180] (Theta);
\draw [->, line width=0.75pt] (X2) to[in=60, out=-10] (Y);

\draw [->, line width=0.75pt] (A_D) to (Theta);

\end{tikzpicture}
\caption{Single-world intervention graph with English language learner (ELL) status $A$, English lexical proficiency $A_Y$, instructional English comprehension $A_\Theta$, potential algebra ability $\Theta^{a_\Theta}$, potential item response $Y^{a_Y, a_\Theta}$, and confounders $W$: socio-economic status (SES) and immigration status (IS). \small{The blue arrow indicates an unfair path, and the bold arrow indicates a deterministic relationship, $A \equiv A_Y \equiv A_\Theta$.}}\label{fig:DAG_bicycle}
\end{figure}

\section{Detecting Separable DIF}

When the aforementioned identification assumptions hold, we estimate the separable DIF effects. In this section, we introduce two methods: one for simple separable DIF and the other for general separable DIF.

\subsection{Detection Method for Simple Separable DIF}

To detect the simple separable DIF effect, $\tau_{\rm SS}(\theta)$, we propose a new method using causal forests. Causal forests extend the random forest algorithm to the estimation of heterogeneous treatment effects, in particular CATE. Briefly, the causal forests estimator can be viewed as a weighted linear regression of a residualized outcome on a residualized regressor, where the weights reflect the contribution of each observation to the target point. This estimator typically uses ``honesty'' (sample splitting) to ensure the asymptotic normality of the estimates. For more details, see  \Textcite{wager2018estimation,athey2019generalized}. 

Our method requires known ability or ability scores reliably estimated via item response theory (IRT) models using anchor items. We first fit a causal forest with the item response $Y$, the treatment variable $A$, and the set of covariates $\{X, \Theta\}$. From this fitted model, we compute the doubly robust (DR) scores $\hat{\tau}^{\rm dr}_i$, for the CATE. Next, we find the best linear projection \parencite[BLP;][]{semenova2021debiased} of the CATE onto ability $\Theta$ by fitting a linear regression model of $\hat{\tau}^{\rm dr}_i$ on $\Theta$. While an intercept and a slope coefficient are used in the baseline regression, one can include higher-order terms of $\Theta$ if a more complex relationship is assumed. The DR scores are constructed using Neyman-orthogonal moments, so the resulting coefficients are insensitive to first-order estimation errors in the nuisance functions (i.e., the propensity score and outcome predictions) \parencite{chernozhukov2018doubleml, semenova2021debiased}. To formally detect separable DIF, we perform a joint Wald test using heteroskedasticity-robust standard errors under the null hypothesis that both regression coefficients (intercept and slope) are equal to zero. Lastly, we obtain the $F$-statistic and corresponding $p$-value to conduct a significance test at a pre-specified level (e.g., $\alpha=0.05$). We summarize the implementation steps for simple separable DIF in Algorithm \ref{algo_CF}.  

\begin{figure}[h]
  \makebox[\linewidth]{%
  \begin{minipage}{\dimexpr\linewidth-6em}
\begin{algorithm}[H]
\caption{\small{Detection for Simple Separable DIF}}
\label{algo_CF}
\begin{algorithmic}[1]
\small
\Require Item response $Y$, treatment/group $A$, ability $\Theta$, confounders $X$.
\Require R packages \texttt{grf} (for causal forests) and \texttt{car} (for Wald test).

\State Define the covariate set $X^+ \gets (X, \Theta)$.
\State Fit a causal forest to estimate the CATE $\hat{\tau}_i$:
       $cf \gets \texttt{causal\_forest(Y=Y, W=A, X=X$^+$)}$.
\State Compute doubly robust scores of CATE estimates: $\hat{\tau}^{\rm dr}_i \gets  \texttt{get\_scores}(cf)$
       
\State Fit the linear regression model $\mathcal{M}_1: \hat{\tau}^{\rm dr}_i = \beta_0 + \beta_1 \Theta_i + \epsilon_i$.

\State Define the joint null hypothesis, $H_0$: $\beta_0 = \beta_1 = 0$.
       
\State Perform the joint Wald test using a heteroscedasticity-consistent (HC) variance-covariance matrix.
       \Statex \hspace{\algorithmicindent} $test \gets \texttt{linearHypothesis}(\mathcal{M}_1, \texttt{c("(Intercept) = 0", "theta = 0")},$
       \Statex \hspace{\algorithmicindent} $ \quad\qquad\ \texttt{white.adjust  = "hc3"})$
\State Extract the $F$-statistic $F_{\text{obs}}$ and $p$-value $p_{\text{value}}$ from the test result, $test$.

\State Reject the null hypothesis if $p_{\text{value}} < \alpha$ (e.g., 0.05).
\Ensure $F_{\text{obs}}$, $p_{\text{value}}$, and the decision.
\end{algorithmic}
\end{algorithm}
  \end{minipage}}
\end{figure}
\raggedbottom

We implement the proposed method in R using the \texttt{grf} package \parencite{grf} for causal forests and the \texttt{car} package \parencite{car} for the Wald test. The R code for this method can be found at \url{https://github.com/youmisuk/separableDIF}. We note that other estimation methods, including BART and double machine learning, can also be used to detect the simple separable DIF effect. Additionally, when the estimated ability is used in the detection procedure, measurement error in the ability estimates may affect the DIF detection results. A possible remedy is to use multiple imputation, in which test-takers' abilities are sampled from their posterior distributions, the proposed method is applied to each imputed dataset, and the resulting $F$ statistics are combined using an appropriate pooling rule to obtain an overall test decision.

\subsection{Detection Method for General Separable DIF}

To detect the general separable DIF effect, $\tau_{\rm GS}(\theta)$, we propose a new method using BART. BART for causal inference \parencite{hill2011bayesian} leverages the BART algorithm \parencite{chipman2010BART} to estimate a single flexible outcome model by regressing the outcome on the treatment and covariates. Briefly, this model is represented as a sum of multiple regression trees. The fitted model is used to compute the CATE by comparing the predictions under different treatment conditions. It also supports easy inference within the Bayesian paradigm. For further details, see  \Textcite{chipman2010BART,hill2011bayesian}.

\begin{figure}[!b]
  \makebox[\linewidth]{%
  \begin{minipage}{\dimexpr\linewidth-6em}
\begin{algorithm}[H]
\caption{\small{Detection for General Separable DIF}}
\label{algo_BART}
\begin{algorithmic}[1]
\small
\Require Item response $Y$, group/treatment indicator $A$, mediator $\Theta$, confounders $W$.
\Require R package \texttt{bartCause}.
\State Define the covariate set $W^{+} \gets (W, \Theta)$.
\State Fit the BART outcome model using \texttt{bartCause}:
\Statex \quad $bart \gets \texttt{bartc(response=Y, treatment=A, confounders=W$^+$, \ldots)}$.
\State Specify an ability grid $\mathcal{G}$, e.g., $\mathcal{G} \gets \texttt{seq}(-3,3,\texttt{by=}0.5)$.
\State \textbf{For} each grid value $g \in \mathcal{G}$ \textbf{do}
\Statex \quad Obtain posterior draws of individual-level CATEs at $\Theta=g$:
\Statex \quad $\hat{\tau}^{(d)}(g) \gets \texttt{predict(}bart,\ \texttt{newdata=data.frame(W=W, Theta=g),\ type="icate")}$.
\Statex \quad Compute density weights for $f(\Theta=g \mid A=a, W)$ (assumed or estimated):
\Statex \quad $w_i(g) \gets f(\Theta=g \mid A=a, W_i)$ for $i=1,\ldots,n$.
\Statex \quad Compute the weighted score under draw $d$:
\Statex \quad $\hat{\tau}^{(d)}_{\rm GS}(g) \gets \dfrac{\sum_{i=1}^n \hat{\tau}^{(d)}_i(g)\, w_i(g)}{\sum_{i=1}^n w_i(g)}$.
\State \textbf{end for}
\State Compute the posterior mean curve $\hat{\tau}_{\rm GS}(g) \gets \frac{1}{D}\sum_{d=1}^D \hat{\tau}^{(d)}_{\rm GS}(g)$ for all $g \in \mathcal{G}$.
\State Construct one-sided and two-sided $p$-values for $\hat{\tau}_{\rm GS}(g)$: 
\Statex $\hat{p}^{one}_g \gets \text{min} \left\{ \frac{1}{D}\sum_{d=1}^D I(\hat{\tau}^{(d)}_{\rm GS}(g) < 0),  \frac{1}{D}\sum_{d=1}^D I(\hat{\tau}^{(d)}_{\rm GS}(g) >0) \right\}$; $\hat{p}_g = 2\hat{p}^{one}_g$
\State Adjust the two-sided $p$-value $\hat{p}_g$ to account for multiple testing.
\State Reject $H_0:\tau_{\rm GS}(\theta)=0$ for all $\theta \in \mathcal{G}$ if any adjusted two-sided $p$-value is less than 0.05.
\Ensure $\hat{\tau}_{\rm GS}(g)$ over $\mathcal{G}$, the adjusted p-values, and the decision.
\end{algorithmic}
\end{algorithm}
  \end{minipage}}
\end{figure}

Our detection method for general separable DIF requires either known ability or estimated ability scores via latent regression IRT models using anchor items. Specifically, we first fit a BART model with the item response $Y$, the treatment variable $A$, and the covariate set $\{W, \Theta\}$. Next, we specify an ability grid $g \in \mathcal{G}$. For each grid value $g$, we obtain posterior draws of CATEs from the fitted BART model and compute density weights from the assumed or estimated distribution. We then compute the weighted effect for each posterior draw to derive the posterior mean curve. Subsequently, we compute $p$-values for $\hat{\tau}_{\rm GS}(g)$ from posterior draws and adjust the two-sided $p$-values using a multiple-testing correction method \parencite[here, the Benjamini-Hochberg correction; ][]{benjamini1995controlling}. Finally, we reject the null hypothesis that $\tau_{\rm GS}(\theta) = 0$ for all $\theta \in \mathcal{G}$ if any adjusted $p$-value is less than 0.05. Algorithm \ref{algo_BART} summarizes the implementation steps for general separable DIF, along with \texttt{bartCause} package \parencite{hill2011bayesian}.

We note that alternative estimation methods, including other machine learning methods, or doubly and triply robust methods, can also be used to detect the general separable DIF effect. Additionally, because the proposed procedure conducts multiple correlated pointwise tests and adjusts their $p$-values, the overall test may be conservative, especially when a dense grid of $\theta$ values is used. To improve power, more computationally intensive approaches, such as parametric bootstrap procedures that account for the dependence among tests, may be considered. Moreover, when ability is estimated rather than known, measurement error in the ability estimates may affect the DIF detection results. This uncertainty can be addressed using multiple imputation.\raggedbottom

\subsection{Simulation Study}\label{sec:simu}

We conducted a simulation study to examine the performance of our proposed detection methods using causal forests (i.e., Algorithm \ref{algo_CF}) and BART (i.e., Algorithm \ref{algo_BART}). We investigated whether they can detect separable DIF effects in a test item under different effect conditions: no effect (e.g., $\tau_{\rm SS}(\theta)=0$, $\tau_{\rm GS}(\theta)=0$), constant effect (e.g., $\tau_{\rm SS}(\theta)\equiv c$, $\tau_{\rm GS}(\theta)\equiv c$), and heterogeneous effect (e.g., $\tau_{\rm SS}(\theta)\not\equiv c$, $\tau_{\rm GS}(\theta)\not\equiv c$).

More specifically, our data-generating models were based on logistic regression with binary $A_Y$ and if applicable, $A_\Theta$, unidimensional $\Theta$, and confounders ($X$ or $W$). Then we generated $A$ so that $A \equiv A_Y$ for simple separable DIF and $A \equiv A_Y \equiv A_\Theta$ for general separable DIF. An item's difficulty parameter was set to a random number drawn from a uniform distribution between 0 and 1, and its discrimination parameter was set to 1. We manipulated the effect conditions by varying the coefficients for the main effect of $A_Y$ and its interaction with $\Theta$. In the no-effect condition, both coefficients were set to zero on the logit scale. In the constant effect condition, the main effect coefficient was set to 0.5 on the logit scale, whereas in the heterogeneous effect condition, the main effect and interaction coefficients were set to 0.5 and 0.25 on the logit scale, respectively. In the simulation study, we also varied the sample size across three conditions: $N=1000$, $2000$, and $4000$. See Supplemental Appendix S3 for more details of the data-generating models. 

Additionally, we used both the true/known ability and the estimated ability scores in the simulation study. To estimate the ability, we fit appropriate IRT models with either 30 or 60 anchor items\footnote{We fit the two-parameter logistic (2PL) IRT models when there is no impact, and when impact is present, we fit latent regression 2PL models, where the treatment variable and confounder(s) are included as predictors. After selecting the appropriate IRT models, we obtain the expected a posteriori (EAP) ability score estimates.}, and then used the ability estimates to detect the presence of a separable DIF effect for one item. We include the simulation results with the true ability below and report those with the estimated ability in Supplemental Appendix \ref{app:simu}. Across all conditions, we evaluated the performance of each detection method by measuring Type-I error and power rates in a test item at an alpha level of 0.05. We repeated the simulation with each condition 500 times. The R code for our data-generating models and simulations is available at \url{https://github.com/youmisuk/separableDIF}.

Table \ref{tab:simu_true} summarizes the simulation results across varying sample sizes based on true/known ability. For simple separable DIF, Algorithm \ref{algo_CF} using causal forests shows a well-controlled Type-I error rate across all sample size conditions, and statistical power increases as the sample size increases under the constant and heterogeneous effect conditions. For general separable DIF, Algorithm \ref{algo_BART} yields a conservative Type-I error rate due to multiple testing adjustment, and power rates similarly increase with larger sample sizes in both the constant and heterogeneous effect conditions. These findings confirm that our proposed methods perform effectively across different simulation scenarios, although Algorithm \ref{algo_BART} exhibits slightly conservative control of Type-I errors. 

\begin{table}[hbt]
    \centering
    \caption{Detection rates under different effects and sample size conditions: True ability}\label{tab:simu_true}
   	\vspace{-2mm}
    \small
    \begin{threeparttable}
\begin{tabular}{l  |ccccccccc}
\hline      
Effect       & \multicolumn{3}{c}{Zero} &  \multicolumn{3}{c}{Constant}  & \multicolumn{3}{c}{Heterogeneous} \\ 
Rejection Rate      & \multicolumn{3}{c}{Type-I error} & \multicolumn{3}{c}{Power} & \multicolumn{3}{c}{Power} \\  
Sample size      & 1000 & 2000 & 4000 & 1000 & 2000 & 4000 & 1000 & 2000 & 4000 \\  
\hline 
Simple Separable DIF  & 0.044 & 0.038  & 0.028 & 0.468 & 0.794 & 0.974 & 0.622 &  0.912 & 0.998  \\ 
General Separable DIF  & 0.002  & 0.004 & 0.002 & 0.126 & 0.338 & 0.780 &  0.218 & 0.662 &  0.974   \\ 
\hline       
\end{tabular}
\vspace{-1mm}
  \begin{tablenotes}[para,flushleft]
    \footnotesize Note: DIF = differential item functioning. Simple separable DIF is detected via Algorithm 1, and general separable DIF is detected via Algorithm 2.  
  \end{tablenotes}
  \end{threeparttable}
\end{table}

Additionally, when using the estimated ability, we obtain similar simulation results with 60 anchor items, but find more divergent results with 30 anchor items due to  increased measurement errors. More details are provided in Supplemental Appendix \ref{app:simu}.

\subsection{Empirical Example}\label{sec:realdata}

We use data from a constructed observational study of U.S.-resident adults conducted by \textcite{keller2025new} to demonstrate the proposed approach. We focus on their mathematics pretest, which contains 12 multiple-choice items with a correct response coded as $1$ and an incorrect response coded as $0$. The construct of interest is general quantitative aptitude. Although this test targets U.S.-resident adults, they may have varying English-language backgrounds, so that some items may function differentially at the same level of quantitative aptitude. For example, Items 8 and 10 are as follows:

\begin{quote}

\textbf{8}. How much would it cost to ride 3.5 miles in a taxicab, if the rate is \$1.40 for the first quarter mile and \$.40 for each additional quarter mile?

(a) \$2.50; (b) \$3.50; (c) \$5.60; (d) \$6.60; 
(e) \$8.00
\vspace{1mm}

\textbf{10}. A wholesale book supplier buys copies of a certain paperback book at \$4.80 per dozen and sells them to book dealers at the rate of 18 for \$9.00. How many books must be sold to make a profit of \$50.00? \\
(a) 100; (b) 200; (c) 300; (d) 400; (e) 500
\end{quote}

These two items express numerical quantities as words---for example, a quarter denotes 1/4 and a dozen denotes 12, where Item 10 is the more text-heavy of the two. Motivated by this variation, our non-manipulable treatment $A$ of interest is English-language background (high versus low), which is ill-defined and consists of multiple components. In particular, construct-irrelevant component $A_Y$ of $A$ is English lexical proficiency (proficient versus non-proficient), where $A \equiv A_Y$. In the data, $A_Y$ is constructed by dichotomizing each subject's vocabulary scores at half of the maximum possible score. Additionally, we do not assume $A$ directly influences $\Theta$, given the target population and construct of interest, though $A$ and $\Theta$ can be associated through common causes. The confounders we adjust for are  household income and whether a subject completed college degree. We estimate math ability by fitting a latent regression 2PL model, using the the remaining 10 items as anchors and the confounders as predictors. Since the observed data contain no missing values, we use the entire sample of 2,200 from \textcite{keller2025new}.

In the absence of impact, the effect of interest is \textit{simple separable DIF}. We assume a data-generating model like that in Figure \ref{fig:DAG_no_impact}, except that the confounders influence $A$, $\Theta$, and $Y$. We use Algorithm \ref{algo_CF} (causal forests) to estimate the simple separable DIF effect. We also compare these results with those from two traditional DIF methods---Mantel--Haenszel test \parencite[denoted as MH;][]{holland1986differential} and logistic regression \parencite[denoted as Logistic;][]{swaminathan1990detecting}---, where $A$ serves as the target group membership. The R code for this analysis is available at the first author's GitHub repository. 

Table \ref{tab:result} summarizes the results for the two math items using both the proposed separable DIF approach and traditional DIF methods. The traditional methods (MH and Logistic) flag both items as exhibiting DIF; in particular they favor high English-language background over low background. These capture only the overall association with the non-manipulable attribute and are therefore limited in identifying the real cause of the differential functioning. In contrast, our proposed approach using Algorithm \ref{algo_CF} adjusts for confounders (income and education) and interprets the effect through a manipulable component, namely $A_Y$. The proposed method provides causal evidence of $A_Y$-driven bias for both Item 8 and Item 10. That is, we confirm a nonzero effect of English lexical proficiency on the responses to these items, conditional on ability. Therefore, we attribute test-takers' differential performance to their English lexical proficiency, under the assumed SWIG and stated assumptions. Based on this empirical evidence, the two items should be revised so as to neutralize the problematic interaction between the examinee's construct-irrelevant, manipulable component and the item content.  Overall, these findings illustrate how the proposed separable DIF framework can identify the real cause of item bias and inform item revision.  

\begin{table}[H]
\centering
\caption{Results of differential item functioning (DIF) analyses for the math pretest items from \Textcite{keller2025new}\label{tab:result}} 
\vspace{-2mm}
\small
\begin{threeparttable}
\begin{tabular}{l|R{1.5cm}C{3cm}|R{1.5cm}C{2cm}|R{1.5cm}C{2cm}}
  \hline
 & \multicolumn{2}{c|}{Algorithm 1}  & \multicolumn{2}{c|}{MH} & \multicolumn{2}{c}{Logistic} \\ 
\multicolumn{1}{c|}{Item} & $p$-value &  & $p$-value & & $p$-value & \\
  \hline
8 & $0.003$ & Separable DIF  & $<0.001$ & DIF & $<0.001$ & DIF\\  
10 & $0.012$ & Separable DIF  & $0.001$ & DIF & $0.004$ & DIF\\
\hline
  \end{tabular}
\vspace{-1mm}
  \begin{tablenotes}[para,flushleft]
    \footnotesize Note: MH = Mantel--Haenszel test; Logistic = logistic regression.
  \end{tablenotes}
  \end{threeparttable}
\end{table}

\section{Discussion and Conclusions}\label{sec:con}

This paper introduces a causal framework for separable DIF effects to identify the ``true'' causes of test unfairness associated with non-manipulable treatments. The framework leverages the concepts of treatment decomposition and separable effects to link a non-manipulable treatment $A$ to intervening variables: a construct-irrelevant component $A_Y$ and a construct-relevant component $A_\Theta$. The construct-irrelevant intervening component can be interpreted as the cause of item bias. To summarize, our separable DIF approach offers five key contributions: (i) proposing using treatment decomposition to transform non-manipulable treatments into intervening variables, (ii) defining meaningful separable estimands in the context of test fairness, specifically simple separable DIF and general separable DIF, (iii) outlining the necessary causal assumptions to identify these separable effects, (iv) providing flexible detection methods using causal forests and BART, and (v) extending the utility of causality in psychometric research and testing practices. 

In traditional DIF analysis, even when non-manipulable group memberships, like gender, are used, psychometricians do not simply interpret the source of DIF as being driven by gender itself. Rather, they heuristically link gender-related DIF to specific item characteristics associated with gender-specific experience and skills \parencite{zumbo2007three}. Our study formalizes this heuristic process. We articulate the underlying assumptions, including the deterministic relations and dismissible component conditions, beyond the usual causal assumptions of consistency, ignorability, and positivity. Specifically, a deterministic relationship is assumed between $A$ and each component $A_Y$ or $A_\Theta$ in the observed data. It is conceptually intuitive to assume such a deterministic relation between non-manipulable treatments and their constituent components through the lens of social constructivism and the ``bundle of sticks'' framework \parencite{sen2016race}. Under a social constructivist paradigm, for example, gender is not merely an inherent or biological trait that causes different social experiences; rather, the social category of gender is itself constituted by a bundle of societal roles, expectations, and culturally specific exposures. Therefore, this deterministic relation is arguably met in DIF contexts. Nevertheless, when researchers argue that such a deterministic relation does not hold and four-arm data are instead observed in practice (i.e., $(A_Y, A_\Theta) \in \{(0,0), (0,1), (1,0), (1,1) \}$), they can re-cast this problem as a multi-categorical treatment setting rather than analyzing two-arm data (i.e., $(A_Y, A_\Theta) \in \{(0,0), (1,1) \}$) in our framework. See  \textcite{park2026separable} for more details on separable effects in four-arm designs. 

In contrast, dismissible component conditions reflect the key causal structure among the treatment, ability, and outcome variables, as well as the assumption of no unmeasured confounding in a future trial with random assignment. Under the FFRCISTG model, these conditions imply partial isolation of the two components:  $A_Y$ partial isolation ($\Theta^{a_Y, a_\Theta}=\Theta^{a_Y', a_\Theta}$) and $A_\Theta$ partial isolation ($Y^{a_Y, a_\Theta}=Y^{a_Y, a_\Theta'}$ if $\Theta^{a_\Theta} = \Theta^{a_\Theta'}$) \parencite{stensrud2021generalized}. As such, the  underlying assumptions for separable DIF are more demanding than those in prior work by \textcite{suk_lyu_2026}, which focuses on manipulable treatments or treats non-manipulable treatments as causal variables. Despite this complexity, our study provides complementary causal tools that psychometricians can widely adopt, given their frequent use of non-manipulable treatments in fairness investigations.\footnote{The connection between causal fairness notions and the statistical fairness notion of DIF is detailed in \textcite{suk_lyu_2026}. Briefly, if the treatment is considered manipulable and causal assumptions are satisfied in the absence of confounders, traditional DIF can be interpreted as a causal effect; therefore, traditional test statistics can be used to evaluate this effect. For more details, see \textcite{suk_lyu_2026}.}      

Additionally, attributing item bias to its underlying real causes---examinees' construct-irrelevant, manipulable constituents (e.g., gender-specific exposure or English lexical proficiency)---leads item developers to design targeted interventions. Traditional DIF methods are primarily exploratory tools that detect associations between a group variable of interest and item performance. While exploratory DIF flags differences in item performance across groups, it only captures the overall effect of group membership on item response; it cannot determine whether these differences are driven by an unfair causal pathway or not.  In the absence of such evidence, test developers may have limited justification for investing the time and expense needed to revise an item. However, if our framework reveals a nonzero separable effect, researchers can more confidently attribute the observed effect to construct-irrelevant component $A_Y$. This provides strong causal evidence for revising the target item to eliminate or mitigate the problematic interaction between the examinee's construct-irrelevant constituent and the item content. Our framework thus formalizes the traditionally heuristic practice of item revision by providing a theoretical justification for how unfair items are identified and revised.

Although this framework is tailored to test unfairness, the core idea of using treatment decomposition to address non-manipulable treatments can be applied to other causal settings. For example, causal decomposition analysis \parencite{vanderweele2014causal, jackson2018decomposition} primarily aims to explain outcome disparities by estimating how much an outcome gap would be reduced if mediator(s) were equalized, rather than by intervening on the group variable (e.g., race). Our principle can extend causal decomposition analysis by decomposing the non-manipulable treatment variable into intervening variables: one component that directly affects the outcome and another that affects the outcome only via the mediator. This approach will investigate outcome disparities by intervening on treatment components rather than mediators, and offer a new perspective for addressing outcome disparities in the social sciences. 

Based on our findings, we provide several suggestions for future research on causal fairness in DIF analysis. First, our method assumes known or reliably estimated ability scores using anchor items. In cases where anchor items are unknown, researchers typically estimate ability scores and DIF effects simultaneously. Future research could investigate detection methods for separable DIF effects when anchor items are unknown, potentially using item purification or regularization techniques, and also attempt to estimate the DIF-adjusted ability score using partial measurement invariance strategies. Second, we utilize causal forests and BART for our detection methods, along with a joint Wald test and multiple-testing correction. However, there are other valid causal methods, such as double machine learning, targeted maximum likelihood estimation with super learner, or triply robust methods, and other consistent tests for heterogeneous treatment effects. Future research could investigate how to adopt these methods or enhance our proposed methods (e.g., by improving the statistical power of our Algorithm 2) to further refine separable DIF detection.
Third, our framework assumes no unmeasured confounding. Future work would explore methods that are robust to unmeasured confounding, such as those using instrumental variables or proximal inference \parencite[e.g.,][]{park2024proximal}, or develop sensitivity analysis techniques in this context. 

Overall, our separable DIF framework promotes interventionist approaches in educational testing to guide tangible interpretations and item revisions based on causal DIF analysis. The proposed framework complements both traditional (association-based) DIF methods and existing causal DIF analyses by identifying the underlying sources of item bias and providing targeted guidance for developing fair items. We believe that this interventionist approach will contribute to creating assessments where every test-taker has a fair opportunity to demonstrate their abilities.

\section*{Acknowledgment}
The authors are grateful to Dan Bolt, Chan Park, and Chenguang Pan for their insightful comments on an earlier version of the manuscript.

\printbibliography

\newpage 

\setcounter{equation}{0}
\setcounter{figure}{0}
\setcounter{table}{0}
\setcounter{page}{1}
\setcounter{section}{0}
\renewcommand{\theequation}{S\arabic{equation}}
\renewcommand{\thefigure}{S\arabic{figure}}
\renewcommand{\thesection}{S\arabic{section}}
\renewcommand{\thetable}{S\arabic{table}}
\renewcommand{\thefigure}{S\arabic{figure}}

\begin{center}
{\Large
\textbf{Supplementary Materials}
}
\end{center}

\setcounter{secnumdepth}{2}

\section{Directed Acyclic Graphs}\label{app:DAG}

In this section, we provide directed acyclic graphs (DAGs) for each SWIG used in the main text.

\begin{figure}[ht]
\centering
    \subcaptionbox{DAG with $A$}{
\begin{tikzpicture}
\tikzset{line width=1pt, outer sep=1pt,
ell/.style={draw,fill=white, inner sep=3pt,
line width=1pt},
swig vsplit={gap=5pt,
inner line width right=0.5pt},
ellempty/.style={fill=white}};
\node[name=A,ell,shape=ellipse] at (0,0){$A$}; 
\node[name=Theta,ell,shape=ellipse] at (2,2){$\Theta$};
\node[name=Y,ell,shape=ellipse] at (4, 0){$Y$};
\node[name=X,ell,  shape=ellipse] at (2,-1.3) {$X$};

\draw [->, line width=0.75pt] (Theta) to (Y);
\draw [draw=blue, ->, line width=0.75pt] (A) to (Y);

\draw [->, line width=0.75pt] (X) to[in=-80, out=190] (A);
\draw [->, line width=0.75pt] (X) to[in=-100, out=-10] (Y);
\end{tikzpicture}}
\hspace{0.5in}
    \subcaptionbox{DAG with $A$ and $A_Y$}{
\begin{tikzpicture}
\tikzset{line width=1pt, outer sep=1pt,
ell/.style={draw,fill=white, inner sep=3pt,
line width=1pt},
swig vsplit={gap=5pt,
inner line width right=0.5pt},
ellempty/.style={fill=white}};
\node[name=A,ell,shape=ellipse] at (0,0){$A$}; 
\node[name=A_Y,ell,shape=ellipse] at (2,0){$A_Y$};
\node[name=Theta,ell,shape=ellipse] at (3,2){$\Theta$};
\node[name=Y,ell,shape=ellipse] at (5, 0){$Y$};
\node[name=X,ell,  shape=ellipse] at (2.5,-1.3) {$X$};

\draw [->, line width=1.75pt] (A) to (A_Y);
\draw [->, line width=0.75pt] (Theta) to (Y);
\draw [draw=blue, ->, line width=0.75pt] (A_Y) to (Y);

\draw [->, line width=0.75pt] (X) to[in=-80, out=190] (A);
\draw [->, line width=0.75pt] (X) to[in=-100, out=-10] (Y);
\end{tikzpicture}}
\caption{DAGs of Figure 3 in the main text, with non-manipulable treatment $A$, ability $\Theta$, item response $Y$, confounders $X$, and manipulable treatment component $A_Y$. \small{The blue arrow indicates an unfair path, and the bold arrow indicates a deterministic relationship, $A\equiv A_Y$.}}
\end{figure}

\begin{figure}[h!]
\centering
    \subcaptionbox{DAG with $A$}{
\begin{tikzpicture}
\tikzset{line width=1pt, outer sep=1pt,
ell/.style={draw,fill=white, inner sep=3pt,
line width=1pt},
swig vsplit={gap=5pt,
inner line width right=0.5pt},
ellempty/.style={fill=white}};
\node[name=A,ell,shape=ellipse] at (0,0){$A$}; 
\node[name=Theta,ell,shape=ellipse] at (2,1.6){$\Theta$};
\node[name=Y,ell,shape=ellipse] at (4, 0){$Y$};
\node[name=X,ell,  shape=ellipse] at (1.5,2.8) {$W$};

\draw [->, line width=0.75pt] (Theta) to (Y);
\draw [draw=blue, ->, line width=0.75pt] (A) to (Y);

\draw [->, line width=0.75pt] (X) to[in=100, out=200] (A);
\draw [->, line width=0.75pt] (X) to[in=100, out=300] (Theta);
\draw [->, line width=0.75pt] (A) to (Theta);
\draw [->, line width=0.75pt] (X) to[in=60, out=-10] (Y);
\end{tikzpicture}}
\hspace{0.5in}
    \subcaptionbox{DAG with $A$, $A_Y$ and $A_\Theta$}{
\begin{tikzpicture}
\tikzset{line width=1pt, outer sep=1pt,
ell/.style={draw,fill=white, inner sep=3pt,
line width=1pt},
swig vsplit={gap=5pt,
inner line width right=0.5pt},
ellempty/.style={fill=white}};
\node[name=A,ell,shape=ellipse] at (0,0){$A$}; 
\node[name=A_Y,ell,shape=ellipse] at (2,0){$A_Y$};
\node[name=A_D,ell,shape=ellipse] at (1.5,0.8){$A_\Theta$};
\node[name=Theta,ell,shape=ellipse] at (3,1.6){$\Theta$};
\node[name=Y,ell,shape=ellipse] at (5, 0){$Y$};
\node[name=X,ell,  shape=ellipse] at (2.5,2.8) {$W$};

\draw [->, line width=1.75pt] (A) to (A_Y);
\draw [->, line width=1.75pt] (A) to (A_D);
\draw [->, line width=0.75pt] (Theta) to (Y);
\draw [draw=blue, ->, line width=0.75pt] (A_Y) to (Y);

\draw [->, line width=0.75pt] (X) to[in=100, out=190] (A);
\draw [->, line width=0.75pt] (X) to[in=100, out=300] (Theta);
\draw [->, line width=0.75pt] (A_D) to (Theta);
\draw [->, line width=0.75pt] (X) to[in=60, out=-10] (Y);
\end{tikzpicture}}
\caption{DAGs of Figure 5 in the main text, with non-manipulable treatment $A$, ability $\Theta$, item response $Y$, covariates $W$, and manipulable treatment components $A_Y$ and $A_\Theta$. \small{The blue arrow indicates an unfair path, and the bold arrow indicates a deterministic relationship, $A \equiv A_Y \equiv A_\Theta$.}}
\end{figure}

\begin{figure}[ht]
\centering
\begin{tikzpicture}
\tikzset{line width=1pt, outer sep=1pt,
ell/.style={draw,fill=white, inner sep=3pt,
line width=1pt},
swig vsplit={gap=5pt,
inner line width right=0.5pt},
ellempty/.style={fill=white}};
\node[name=A,ell,shape=ellipse] at (0,0) [label = below:{\footnotesize Gender} ] {$A$}; 
\node[name=A_Y,ell,shape=ellipse] at (2.6,0) [label = above:{\footnotesize Gender-specific exposure} ] {$A_Y$};
\node[name=Theta,ell,shape=ellipse] at (5,2) [label = above:{\footnotesize Algebra ability} ] {$\Theta$};
\node[name=Y,ell,shape=ellipse] at (7, 0) [label = below:{\footnotesize Item response} ] {$Y$};

\draw [->, line width=1.75pt] (A) to (A_Y);
\draw [->, line width=0.75pt] (Theta) to (Y);
\draw [draw=blue, ->, line width=0.75pt] (A_Y) to (Y);
\end{tikzpicture}
\caption{DAG of Figure 7 in the main text, with gender $A$, gender-specific exposure $A_Y$, algebra ability $\Theta$, and item response $Y$. \small{The blue arrow indicates an unfair path, and the bold arrow indicates a deterministic relationship, $A\equiv A_Y$.}}
\end{figure}

\begin{figure}[ht]
\centering
\begin{tikzpicture}
\tikzset{line width=1pt, outer sep=1pt,
ell/.style={draw,fill=white, inner sep=3pt,
line width=1pt},
swig vsplit={gap=5pt,
inner line width right=0.5pt},
ellempty/.style={fill=white}};
\node[name=A,ell,shape=ellipse] at (0,0) [label = below:{\footnotesize ELL} ] {$A$}; 
\node[name=A_Y,ell,shape=ellipse] at (2.6,0) [label = below:{\footnotesize {\shortstack{Lexical\\proficiency}}} ] {$A_Y$};
\node[name=A_D,ell,shape=ellipse] at (2.3,1) [label = above:{\footnotesize {\shortstack{Instructional\\comprehension}}} ] {$A_\Theta$};
\node[name=Theta,ell,shape=ellipse] at (4.5,2) [label = below:{\footnotesize {\shortstack{Algebra\\ability}}} ] {$\Theta$};
\node[name=Y,ell,shape=ellipse] at (7, 0) [label = below:{\footnotesize Item response} ] {$Y$};
\node[name=X,ell,  shape=ellipse] at (4,3.5) {IS};
\node[name=X2,ell,  shape=ellipse] at (6.5,3) {SES};

\draw [->, line width=1.75pt] (A) to (A_Y);
\draw [->, line width=1.75pt] (A) to (A_D);
\draw [->, line width=0.75pt] (Theta) to (Y);
\draw [draw=blue, ->, line width=0.75pt] (A_Y) to (Y);

\draw [->, line width=0.75pt] (X) to[in=100, out=190] (A);
\draw [->, line width=0.75pt] (X) to[in=110, out=-10] (Y);

\draw [->, line width=0.75pt] (X2) to[in=60, out=180] (Theta);
\draw [->, line width=0.75pt] (X2) to[in=60, out=-10] (Y);

\draw [->, line width=0.75pt] (A_D) to (Theta);

\end{tikzpicture}
\caption{DAG of Figure 8 in the main text, with English language learner (ELL) status $A$, English lexical proficiency $A_Y$, Instructional English comprehension $A_\Theta$, algebra ability $\Theta$, item response $Y$, and confounders $W$: socio-economic status (SES) and immigration status (IS). \small{The blue arrow indicates an unfair path, and the bold arrows indicate a deterministic relationship, $A \equiv A_Y \equiv A_\Theta$.}}
\end{figure}

\vspace{\fill}\clearpage

\section{Identification}\label{app:iden}

\subsection{Simple Separable DIF}

We identify $E(Y^{a_Y=a} \mid \Theta=\theta, X)$ as follows:

\begin{align*}
E(Y \mid A=a,\Theta=\theta,X) 
&= E(Y \mid A_{Y}=a,\Theta=\theta,X) && (\text{Determinism}) \\
&= E(Y^{a_{Y}=a} \mid A_{Y}=a,\Theta=\theta,X) && (\text{A1}) \\
&= E(Y^{a_{Y}=a} \mid \Theta=\theta,X), && (\text{A2 and A3}) 
\end{align*}
and we identify $E(Y^{a_Y=a} \mid \Theta=\theta)$ as:
\begin{align*}
E(Y^{a_Y=a} \mid \Theta=\theta) = E\{E(Y \mid A=a,\Theta=\theta,X)\mid \Theta=\theta\}.
\end{align*}

Therefore, the simple separable DIF is identified by:
\begin{align*}
\tau_{\rm SS}(\theta) &\coloneqq E(Y^{a_Y=a} \mid \Theta=\theta) - E(Y^{a_Y=a'} \mid \Theta=\theta) \\
&= E\{ E( Y \mid A=a, \Theta=\theta, X ) \mid \Theta = \theta\}
- E\{ E( Y \mid A=a', \Theta=\theta, X ) \mid \Theta = \theta\}.
\end{align*}
Note that we still require Assumption (A4) to interpret this effect as the effect of the manipulable intervening component $A_Y$. Without (A4), the same adjustment formula identifies a purely statistical contrast in conditional means of $Y$, but its interpretation as a component-specific effect would not be justified. \raggedbottom

\subsection{General Separable DIF}

The outcome $Y$ is binary, and the ability $\Theta$ is typically continuous. A counterfactual expectation of interest is: 
\begin{align}
E(Y^{a_Y=a', a_\Theta=a} \mid \Theta^{ a_\Theta=a}=\theta ) &= P(Y^{a_Y=a', a_\Theta=a} = 1 \mid \Theta^{ a_\Theta=a}=\theta ) \nonumber \\
&= \frac{ f(Y^{a_Y=a', a_\Theta=a} = 1, \Theta^{ a_\Theta=a} = \theta)}{f(\Theta^{ a_\Theta=a} = \theta)}. \label{eq:GSone}
\end{align}

The joint density $f(Y^{a_{Y}=a',a_{\Theta}=a}=y,\Theta^{a_{\Theta}=a}=\theta,W)$ can be expressed using the chain rule as:
\begin{align}\label{eq:fulljoint}
f(Y^{a_{Y}=a',a_{\Theta}=a}=y,\Theta^{a_{\Theta}=a}=\theta,W) = f(Y^{a_{Y}=a',a_{\Theta}=a}=y \mid \Theta^{a_{\Theta}=a}=\theta,W)f(\Theta^{a_{\Theta}=a}=\theta \mid W)f(W)
\end{align}

Under Assumptions (B1)--(B4) and determinism condition (i.e., $A \equiv A_Y \equiv A_\Theta$) in the observed world, the first term on the right-hand side is identified as follows:
\begin{align*}
f(Y=y \mid A=a',\Theta=\theta,W) 
&= f(Y=y \mid A_{Y}=a',A_{\Theta}=a',\Theta=\theta,W) && (\text{Determinism}) \\
&= f(Y=y \mid A_{Y}=a',A_{\Theta}=a,\Theta=\theta,W) && (\text{B4(i)}) \\
&= f(Y^{a_{Y}=a',a_{\Theta}=a}=y \mid A_{Y}=a',A_{\Theta}=a,\Theta^{a_{\Theta}=a}=\theta,W) && (\text{B1}) \\
&= f(Y^{a_{Y}=a',a_{\Theta}=a}=y \mid \Theta^{a_{\Theta}=a}=\theta,W). && (\text{B2 and B3})
\end{align*}
Similarly, the second term is identified by:
\begin{align*}
f(\Theta=\theta \mid A=a,W) 
&= f(\Theta=\theta \mid A_{Y}=a,A_{\Theta}=a,W) && (\text{Determinism}) \\
&= f(\Theta=\theta \mid A_{Y}=a',A_{\Theta}=a,W) &&  (\text{B4(ii)}) \\
&= f(\Theta^{a_{\Theta}=a}=\theta \mid A_{Y}=a',A_{\Theta}=a,W) &&  (\text{B1}) \\
&= f(\Theta^{a_{\Theta}=a}=\theta \mid W). &&  (\text{B2 and B3})
\end{align*}
Therefore, the joint density \eqref{eq:fulljoint} is identified as: 
\begin{align*}
f(Y^{a_{Y}=a',a_{\Theta}=a}=y,\Theta^{a_{\Theta}=a}=\theta,W) = f(Y=y \mid A=a',\Theta=\theta,W)f(\Theta=\theta \mid A=a,W)f(W).
\end{align*}

Using this result, we identify the counterfactual expectation \eqref{eq:GSone}. Specifically, the numerator is:
\begin{align*}
f(Y^{a_Y=a', a_\Theta=a} = 1, \Theta^{ a_\Theta=a} = \theta) &= \sum_{w}f(Y^{a_{Y}=a',a_{\Theta}=a}=1,\Theta^{a_{\Theta}=a}=\theta,W=w) \\
&= \sum_{w}f(Y=1 \mid A=a',\Theta=\theta,W=w)f(\Theta=\theta \mid A=a,W=w)f(W=w) \\
&= \sum_{w}E(Y \mid A=a',\Theta=\theta,W=w)f(\Theta=\theta \mid A=a,W=w)f(W=w) \\
&= E_{W}[E(Y \mid A=a',\Theta=\theta,W)f(\Theta=\theta \mid A=a,W)].
\end{align*}
The denominator is:
\begin{align*}
f(\Theta^{ a_\Theta=a} = \theta) &= \sum_w f(\Theta = \theta  \mid A=a, W=w) f(W=w) \\ 
&=E_{W}[f(\Theta=\theta \mid A=a,W)].
\end{align*}
Therefore, the general separable DIF is identified by:
\begin{align*}
\tau_{\rm GS}(\theta) &\coloneq E[Y^{a_{Y}=a,a_{\Theta}=a} \mid \Theta^{a_{\Theta}=a}=\theta] - E[Y^{a_{Y}=a',a_{\Theta}=a} \mid \Theta^{a_{\Theta}=a}=\theta] \\
&= \frac{E_{W} [\{E(Y  \mid  A = a , \Theta = \theta, W) - E(Y  \mid  A = a', \Theta = \theta, W)\} f(\Theta = \theta  \mid  A = a, W)]}{E_{W} [f(\Theta = \theta  \mid  A = a, W)]}
\end{align*}

\newpage

\section{Data-Generating Processes in Simulations}\label{app:DGP}

The details of the data-generating processes (DGPs) in our simulation study for \textit{simple separable DIF} are provided below.

\begin{enumerate}

\item Set the sample size to $n$ individuals, indexed by $i=1,\ldots, n$.

\item Generate a confounder for each individual: $X_i \sim \mathcal{U}(-1, 1)$.

\item Generate the individual treatment status $A$ and the treatment component $A_Y$, both of which are affected by the confounder: 
\begin{align*}
A_i \sim Bernoulli(1/(1+e^{-0.5X_i})), \ \ A_{Y,i} \sim Bernoulli(1/(1+e^{-0.8X_i}))
\end{align*}

\item Generate an ability score: $\Theta_i \sim \mathcal{N}(0, 1)$.

\item Generate the observed outcomes for $J$ anchor items ($j=1, \ldots, J$) using the following logistic regression model, where $b_j$ represents the item difficulty parameter:
\begin{align*}
logit(\pi_{ij}) &= \Theta_i  - b_j,  \quad{} \ b_j \sim \mathcal{U}(0, 1), \\
Y_{ij} &\sim Bernoulli(\pi_{ij})
\end{align*}

\item Generate the potential outcomes for the DIF item ($j=J+1$) using the following logistic regression model:
\begin{align*}
logit(\pi_{i, J+1}^{a_Y}) &= \Theta_i  - b_j - (\gamma_1 + \gamma_2 \Theta_i) a_Y + X_i,  \quad{} \ b_j \sim \mathcal{U}(0, 1), \\
Y_{i, J+1}^{a_Y} &\sim Bernoulli(\pi_{i, J+1}^{a_Y})
\end{align*}
Here, $\gamma_1$ and $\gamma_2$ control the form of the DIF effect. To represent a zero DIF effect, $\gamma_1=0$ and $\gamma_2=0$;  for a constant DIF effect, $\gamma_1=1$ and  $\gamma_2=0$;  for a heterogeneous DIF effect, $\gamma_1=1$ and $\gamma_2=0.5$.

\item Generate the observed outcome for the DIF item: 
\[
Y_{i, J+1} = Y_{i, J+1}^{1}A_{Y,i} + Y_{i, J+1}^{0}(1-A_{Y,i}). 
\]

\item Construct the observed dataset $(X, \Theta, Y, A_Y)$ for all  items for the subgroup of $A_i=A_{Y,i}$.

\end{enumerate}

We also provide the details of the DGPs for \textit{general separable DIF} below.

\begin{enumerate}

\item Set the sample size to $n$ individuals, indexed by $i=1,\ldots, n$.

\item Generate a confounder for each individual: $X_i \sim \mathcal{U}(-1, 1)$.

\item Generate individual treatment status $A$ and treatment component $A_Y$, both of which are affected by the confounder: 
\begin{align*}
A_{\Theta,i} \sim Bernoulli(1/(1+e^{-0.5X_i})), \ \ A_{Y,i} \sim Bernoulli(1/(1+e^{-0.8X_i}))
\end{align*}

\item Generate a potential ability score that is affected by the confounder and $A_\Theta$: $\Theta_i^{a_\Theta} \sim \mathcal{N}(X_i + 0.5a_\Theta, 1)$. 

\item Generate the observed outcomes for $J$ anchor items f ($j=1, \ldots, J$) using the following logistic regression model, where $b_j$ represents the item difficulty parameter:
\begin{align*}
logit(\pi_{ij}) &= \Theta_i  - b_j,  \quad{} \ b_j \sim \mathcal{U}(0, 1), \\
Y_{ij} &\sim Bernoulli(\pi_{ij})
\end{align*}

\item Generate the potential outcomes for the DIF item ($j=J+1$) using the following logistic regression model:
\begin{align*}
logit(\pi_{i, J+1}^{a_Y, a_\Theta}) &= \Theta_i^{a_\Theta}  - b_j - (\gamma_1 + \gamma_2 \Theta_i^{a_\Theta}) a_Y + X_i,  \quad{} \ b_j \sim \mathcal{U}(0, 1), \\
Y_{i, J+1}^{a_Y, a_\Theta} &\sim Bernoulli(\pi_{i, J+1}^{a_Y, a_\Theta})
\end{align*}
Here, $\gamma_1$ and $\gamma_2$ control the form of the DIF effect. To represent a zero DIF effect, $\gamma_1=0$ and $\gamma_2=0$;  for a constant DIF effect, $\gamma_1=1$ and  $\gamma_2=0$;  for a heterogeneous DIF effect, $\gamma_1=1$ and $\gamma_2=0.5$.

\item Generate the observed ability and outcome for the DIF item: 
\begin{align*}
\Theta_{i} &= \Theta_{i}^{1}A_{\Theta,i} + \Theta_{i}^{0}(1-A_{\Theta,i}), \\ 
Y_{i, J+1} &= Y_{i, J+1}^{1,1}\mathbbm{1}(A_{Y,i}=1, A_{\Theta,i}=1) + Y_{i, J+1}^{1,0}\mathbbm{1}(A_{Y,i}=1, A_{\Theta,i}=0) + \\
 & \quad{} Y_{i, J+1}^{0,1}\mathbbm{1}(A_{Y,i}=0, A_{\Theta,i}=1) + Y_{i, J+1}^{0,0}\mathbbm{1}(A_{Y,i}=0, A_{\Theta,i}=0). 
\end{align*}

\item Construct the observed dataset $(X, \Theta, Y, A_Y, A_\Theta)$ for all items for the subgroup of $A_{Y,i}=A_{\Theta,i}$.

\end{enumerate}

\newpage

\section{Simulations with Estimated Ability}\label{app:simu}

In this section, we provide the simulation results using the estimated ability scores from latent regression IRT models. We varied the sample size and the number of anchor items. We find results similar to those with the true ability when the number of anchor items is 60 within each sample size condition. More specifically,  for simple separable DIF, Type-I error rates are well-controlled, and statistical power exceeds 0.9 as the sample size reaches 4000. For general separable DIF, Type-I error rates are conservative, while power rates increase as the sample size grows.

When the number of anchor items is reduced to 30, the estimated ability contains greater measurement errors. Thus, the results using these ability estimates show more divergence from those using true ability. However, the absolute differences are below 0.04, except for the power rates for general separable DIF under the two non-zero effect conditions.

\begin{table}[hbt]
    \centering
    \caption{Detection rates under different effects and sample size conditions: Estimated ability}\label{tab:simu_estimated}
   	\vspace{-2mm} 
    \small
    \begin{threeparttable}
\begin{tabular}{l  |ccccccccc}
\hline      
Effect       & \multicolumn{3}{c}{Zero} &  \multicolumn{3}{c}{Constant}  & \multicolumn{3}{c}{Heterogeneous} \\ 
Rejection Rate      & \multicolumn{3}{c}{Type-I error} & \multicolumn{3}{c}{Power} & \multicolumn{3}{c}{Power} \\  
Sample size      & 1000 & 2000 & 4000 & 1000 & 2000 & 4000 & 1000 & 2000 & 4000 \\  
\hline 
\textit{Num. Anchor = 60} \\
Simple Separable DIF  & 0.044 & 0.048 & 0.034 & 0.468 &  0.786 & 0.972 & 0.622 & 0.892 & 0.998  \\ 
General Separable DIF  & 0.000  & 0.004 & 0.001 & 0.108  & 0.328   & 0.734 & 0.206 & 0.590 &  0.970   \\ 
\textit{Num. Anchor = 30} \\
Simple Separable DIF  & 0.066  & 0.074 & 0.046 & 0.450 & 0.794  & 0.978 & 0.622 & 0.920 & 0.994  \\ 
General Separable DIF  &  0.006  & 0.004  & 0.006  & 0.084  & 0.272 & 0.692  &  0.230 & 0.630 &  0.966   \\ 
\hline       
\end{tabular}
\vspace{-1mm}
  \begin{tablenotes}[para,flushleft]
    \footnotesize Note: DIF = differential item functioning. Simple separable DIF is detected via Algorithm 1, and general separable DIF is detected via Algorithm 2.  
  \end{tablenotes}
  \end{threeparttable}
\end{table}

\end{document}